\documentclass[12pt]{iopart}
\usepackage{graphicx}
\usepackage{blindtext}

\expandafter\let\csname equation*\endcsname\relax
\expandafter\let\csname endequation*\endcsname\relax
\usepackage{amsmath,amsthm,amsfonts,amssymb,amscd}
\usepackage{iopams}
\usepackage{verbatim}
\usepackage{mathrsfs}
\usepackage[dvipsnames]{xcolor}
\usepackage{calc}
\usepackage{pgfplots}
\usepackage{listings}
\usepackage{tikz}
\usetikzlibrary{decorations.text}
\usepackage{physics}
\usepackage{epigraph}
\usepackage{cite}
\usepackage{float}
\usepackage{mathtools}
\usepackage{caption}
\usepackage{subcaption}
\usepackage{bm}
\usepackage[colorlinks=true,linkcolor=blue,allcolors=blue,citecolor=blue]{hyperref}
\usepackage[capitalise]{cleveref}
\usepackage[export]{adjustbox}
\newcommand{\pd}{\partial}
\newcommand{\ep}{\epsilon}
\newcommand{\wrt}{\mathrm{d}}
\newcommand{\config}{\mathcal{C}}
\newcommand{\dr}{\wrt \textbf{r}}
\newcommand{\prob}{\mathcal{P}}
\newcommand{\ham}{\mathcal{H}}
\newcommand{\rar}{\rightarrow}

\newcommand{\nh}{\hat{n}}
\newcommand{\rh}{\hat{\rho}}
\newcommand{\ci}{\mathrm{i}}

\begin{document}
\title{Hydrodynamic equations and near critical large deviations of active lattice gases}

\author{Luke Neville}

\address{School of Mathematics, Fry Building, University of Bristol, BS8 1UG, United Kingdom}

\ead{luke.neville@bristol.ac.uk}

\begin{abstract}
Using a path integral approach, we derive and study the hydrodynamic equations and large deviation functions for three active lattice gases. After a review of the path integral for master equations, we first look at a one dimensional model of motility induced phase separation (MIPS), re-deriving the large deviation function that was previously found through a mapping to the ABC model. After extracting the deterministic hydrodynamic equations from the large deviation function, we analyse them perturbatively near the MIPS critical point using a weakly non-linear analysis. Doing this we show that they reduce to equilibrium Model B very close to criticality, with non-equilibrium, or Active Model B terms emerging as we leave the critical region. The same type of weakly non-linear analysis is then applied to the full large deviation function, and we show that the near critical stationary probability distribution is given by the exponential of a $\phi^4$ free energy, as expected in ordinary equilibrium phase separation. Similar calculations are then done for the two other lattice gases one which is another MIPS model, and another which models flocking, and in both cases we find analogous results.

\end{abstract}

\section{Introduction}
Motile active matter consists of self-propelling particles that keep moving by constantly using energy to perform work \cite{ramaswamy2010mechanics,marchetti2013hydrodynamics,cates2019active,gompper20202020}.  
This breaking of detailed balance mean that active matter can show behaviour that equilibrium thermodynamics would forbid, like phase separation in the presence of repulsive interactions \cite{cates2015motility,tailleur2008statistical,fily2012athermal}, or the formation of flocks in two dimensions \cite{toner1995long,vicsek1995novel}. Although originally discovered in agent based simulations \cite{vicsek1995novel}, these effects are now best understood through the use of continuous hydrodynamic equations  \cite{marchetti2013hydrodynamics,cates2019active,stenhammar2013continuum,catesbinary}. 
Similar to equilibrium, this approach uses symmetry to pick out the hydrodynamic variables, before phenomenologically constructing dynamical equations governing their slow evolution in space and time \cite{chaikin1995principles,halperin2019hohenberg,martin1972unified,toner1995long}. As the resulting equations are ambivalent to microscopic detail they have can give insight into universal properties like scaling exponents \cite{tjhung2018cluster,toner2005hydrodynamics,chate2020dry}.

Despite this, it has always been desirable to connect these active hydrodynamic theories to microscopic models through coarse graining. The reason for this is two-fold. One, it helps to understand the range of validity of these equations. For example, are these hydrodynamic equations valid everywhere, or  perhaps just near a critical point where the renormalisation group justifies an expansion in fields and gradients \cite{chaikin1995principles,tauber2014critical}. And two, it gives information about the transport coefficients, telling us, say, whether they should be positive of negative. In the case of active, or motility induced phase separation (MIPS), this connection has been achieved for self propelling particles interacting through quorum sensing \cite{stenhammar2013continuum,solon2015comparative}, with the hydrodynamics matching the microscopic simulations quantitatively well, even being able to predict the binodals, or phase separated densities \cite{solon2018generalized,dinelli2024fluctuating}. Although discovered first, the situation is somewhat worse for flocking, where, although the hydrodynamic equations can be derived \cite{bertin2009hydrodynamic,bertin2013mesoscopic,chate2020dry,te2023derive,degond2008continuum}, the predicted parameter values do not quantitatively match particle based simulations. The reason for this discrepancy comes from the number of uncontrolled approximations made in the derivation \cite{te2023derive,dinelli2024fluctuating}. For example, many of these derivations rely on a truncation of an infinite hierarchy of PDEs, an approximation that cannot be justified a priori \cite{chate2020dry}.

Given these difficulties it seems necessary to find  models that can be coarse grained in a mathematically controlled way, and a class that seems to fill this requirement is the diffusive lattice gas \cite{kipnis2013scaling,spohn2012large,derrida2011microscopic}. These stochastic models originated in the mathematical physics literature \cite{kipnis1982heat,de1985rigorous,demasi1986reaction,kipnis1989hydrodynamics},  and although often applied to boundary driven equilibrium systems \cite{derrida1992exact}, they have recently been applied to \textit{internally} driven active matter \cite{thompson2011lattice,solon2013revisiting,solon2015flocking,kourbane2018exact}.  
Although perhaps more artificial than traditional particle based simulations, they are valuable because their microscopic dynamics means that they become effectively deterministic in the limit of small lattice spacing, with their average behaviour following a set `exact' hydrodynamic equations. As such, these diffusive lattice gases can undergo phase transitions even in one dimension \cite{de1985rigorous,de1993physics,clincy2003phase}, something which is not possible in ordinary one-dimensional models due to the excess of fluctuations \cite{kardar2007statisticalfields,goldenfeld2018lectures}. In 2018, two such lattice gases were introduced by Kourbane-Housenne et al. \cite{kourbane2018exact}, one of which modelled MIPS, the other flocking, and their deterministic hydrodynamic equations were found rigorously \cite{erignoux2021hydrodynamic}. Simulations of the hydrodynamic equations and underlying lattice gas were then compared, and were shown to be in near perfect agreement.

These deterministic hydrodynamic equations capture the most probable behaviour, however the underlying stochasticity means that trajectories can deviate wildly from the deterministic hydrodynamics. Such rare events, or \textit{large deviations} may also be examined, and a generalisation of the central limit theorem tells us that they are exponentially unlikely \cite{jack2010large,jack2020ergodicity,touchette2009large,touchette2011basic}. More precisely, these trajectories have probability
\begin{equation}
\prob[\text{traj}]\asymp \exp[-\mathcal{A}/\ep],
\end{equation}
where $\ep$ is a small parameter usually related to the lattice spacing. The large deviation function (LDF), $\mathcal{A}$, controls the probability of a given trajectory, and so is analogous to the standard action in dynamic field theory \cite{bertini2001fluctuations,bertini2002macroscopic,bertini2007stochastic,bertini2015macroscopic,derrida2011microscopic,bodineau2005distribution,bodineau2008long,bodineau2010current}. 

In this paper we use these methods to take the two lattice models of active matter introduced in Ref. \cite{kourbane2018exact} and connect them with other, more standard hydrodynamic theories of active matter \cite{cates2019active}. For, say, the model of MIPS, this is required because the hydrodynamic equations derive in Ref. \cite{kourbane2018exact} govern two fields, the density and the magnetisation, whereas standard field theories of MIPS like active model B use only one \cite{tjhung2018cluster}. These two fields appear naturally in the derivation of the hydrodynamic equations from the lattice gas, coming from the two types of particle in the lattice gas. In usual studies of MIPS it is argued that all fields apart from the density may be adiabatically eliminated \cite{dinelli2024fluctuating}, however for the lattice gas hydrodynamics the magnetisation evolves on the same time scale and so cannot. In this work we take inspiration from standard gradient expansions \cite{solon2018generalized,speck2015dynamical} to show that this statement, although true in general, does not hold near the MIPS critical point where only the density is critically slowed. As such we are able to apply a weakly non-linear analysis to show that the full lattice gas hydrodynamics can be mapped onto active model B near the critical point. In fact, we show to leading order in a reduced temperature type parameter, that  the dynamics become perfectly equilibrium-like near the critical point, with non-equilibrium terms emerging as one leaves the critical region. After doing this calculation on the deterministic hydrodynamic equations, we then include fluctuations by applying the same weakly non-linear theory to the full LDF. This allows us to show that the near critical stationary probability distribution is the exponential of a $\phi^4$ free energy, just as in equilibrium phase separation. We then perform the same calculations on a new lattice model of MIPS based on quorum sensing, and the model of flocking introduced in Ref. \cite{kourbane2018exact}, finding analogous results in both cases.

The rest of the paper is structured as follows: in Sec. \ref{sec:framework} we introduce lattice gases, the field theoretic framework, and show how to take proper hydrodynamic limits. In Sec. \ref{sec:MIPS} we study the first model of MIPS, re-deriving the known LDF and analysing its near critical behaviour. We then repeat the same calculation for a new model of MIPS based on quorum sensing in Sec. \ref{sec:quorum}, before turning to the flocking model in Sec. \ref{sec: flocking}. Our results are then discussed in Sec. \ref{sec:discussion}. 

\section{Lattice gases and Path integrals}
\label{sec:framework}
Here we introduce stochastic lattice gases and show how to derive their hydrodynamic large deviation functions using  path integrals.

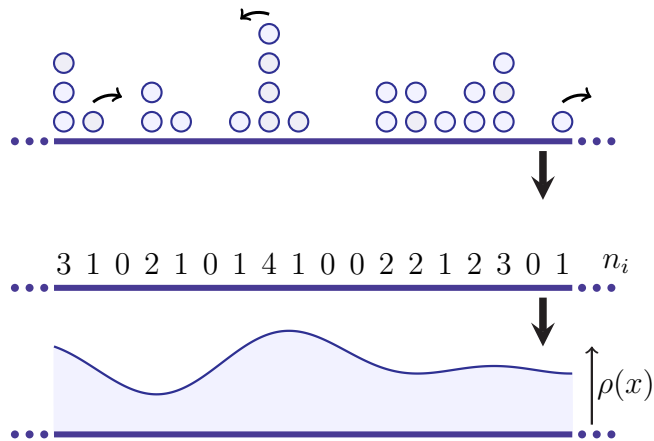
\begin{figure}[t]
    \centering
    \begin{tikzpicture}[scale=0.65]
\draw[-,color=BlueViolet,line width= .8mm] (0,0) -- (10.6,0);
\filldraw[color=BlueViolet, fill=BlueViolet,thick] (-0.2,0) circle (.06);
\filldraw[color=BlueViolet, fill=BlueViolet,thick] (-0.5,0) circle (.06);
\filldraw[color=BlueViolet, fill=BlueViolet,thick] (-0.8,0) circle (.06);
\filldraw[color=BlueViolet, fill=BlueViolet,thick] (10.8,0) circle (.06);
\filldraw[color=BlueViolet, fill=BlueViolet,thick] (11.1,0) circle (.06);
\filldraw[color=BlueViolet, fill=BlueViolet,thick] (11.4,0) circle (.06);
\filldraw[color=Blue, fill=blue!5,thick] (0.2,.4) circle (.2);
\filldraw[color=Blue, fill=blue!5,thick] (0.2,1) circle (.2);
\filldraw[color=Blue, fill=Blue!5,thick] (0.2,1.6) circle (.2);
\filldraw[color=Blue, fill=Blue!5,thick] (0.8,.4) circle (.2);
\filldraw[color=Blue,fill=blue!5,thick] (2,.4) circle (.2);
\filldraw[color=Blue, fill=blue!5,thick] (2,1) circle (.2);
\filldraw[color=Blue, fill=blue!5,thick] (2.6,.4) circle (.2);
\filldraw[color=Blue, fill=blue!5,thick] (3.8,.4) circle (.2);
\filldraw[color=Blue, fill=Blue!5,thick] (4.4,.4) circle (.2);
\filldraw[color=Blue, fill=Blue!5,thick] (4.4,1) circle (.2);
\filldraw[color=Blue, fill=Blue!5,thick] (4.4,1.6) circle (.2);
\filldraw[color=Blue, fill=blue!5,thick] (4.4,2.2) circle (.2);
\filldraw[color=Blue, fill=Blue!5,thick] (5,.4) circle (.2);
\filldraw[color=Blue, fill=blue!5,thick] (6.8,.4) circle (.2);
\filldraw[color=Blue, fill=blue!5,thick] (6.8,1) circle (.2);
\filldraw[color=Blue, fill=Blue!5,thick] (7.4,.4) circle (.2);
\filldraw[color=Blue, fill=blue!5,thick] (7.4,1) circle (.2);
\filldraw[color=Blue, fill=blue!5,thick] (8,.4) circle (.2);
\filldraw[color=Blue, fill=blue!5,thick] (8.6,.4) circle (.2);
\filldraw[color=Blue, fill=blue!5,thick] (8.6,1) circle (.2);
\filldraw[color=Blue, fill=blue!5,thick] (9.2,.4) circle (.2);
\filldraw[color=Blue, fill=Blue!5,thick] (9.2,1) circle (.2);
\filldraw[color=Blue, fill=blue!5,thick] (9.2,1.6) circle (.2);
\filldraw[color=Blue, fill=blue!5,thick] (10.4,.4) circle (.2);
\draw[-to,color=black, line width=.4mm] (10.4,.8) arc (130:70:.6);
\draw[-to,color=black, line width=.4mm] (.8,.8) arc (130:70:.6);
\draw[-to,color=black, line width=.4mm] (4.4,2.6) arc (70:130:.6);
\draw[-,color=BlueViolet,line width= .8mm] (0,-3) -- (10.6,-3);
\filldraw[color=BlueViolet, fill=BlueViolet,thick] (-0.2,-3) circle (.06);
\filldraw[color=BlueViolet, fill=BlueViolet,thick] (-0.5,-3) circle (.06);
\filldraw[color=BlueViolet, fill=BlueViolet,thick] (-0.8,-3) circle (.06);
\filldraw[color=BlueViolet, fill=BlueViolet,thick] (10.8,-3) circle (.06);
\filldraw[color=BlueViolet, fill=BlueViolet,thick] (11.1,-3) circle (.06);
\filldraw[color=BlueViolet, fill=BlueViolet,thick] (11.4,-3) circle (.06);
\node at (.2,-2.5){$3$};
\node at (.8,-2.5){$1$};
\node at (1.4,-2.5){$0$};
\node at (2,-2.5){$2$};
\node at (2.6,-2.5){$1$};
\node at (3.2,-2.5){$0$};
\node at (3.8,-2.5){$1$};
\node at (4.4,-2.5){$4$};
\node at (5,-2.5){$1$};
\node at (5.6,-2.5){$0$};
\node at (6.2,-2.5){$0$};
\node at (6.8,-2.5){$2$};
\node at (7.4,-2.5){$2$};
\node at (8,-2.5){$1$};
\node at (8.6,-2.5){$2$};
\node at (9.2,-2.5){$3$};
\node at (9.8,-2.5){$0$};
\node at (10.4,-2.5){$1$};

\filldraw[color=BlueViolet, fill=BlueViolet,thick] (-0.2,-6) circle (.06);
\filldraw[color=BlueViolet, fill=BlueViolet,thick] (-0.5,-6) circle (.06);
\filldraw[color=BlueViolet, fill=BlueViolet,thick] (-0.8,-6) circle (.06);
\filldraw[color=BlueViolet, fill=BlueViolet,thick] (10.8,-6) circle (.06);
\filldraw[color=BlueViolet, fill=BlueViolet,thick] (11.1,-6) circle (.06);
\filldraw[color=BlueViolet, fill=BlueViolet,thick] (11.4,-6) circle (.06);
\draw[-stealth,color=Black,line width=1mm] (10,-0.2) -- (10,-1.2);
\draw[-stealth,color=Black,line width=1mm] (10,-3.2) -- (10,-4.2);
\draw [color=blue!5,fill=blue!5, smooth, samples=100, domain=0:10.6] plot(\x, {cos(\x*80)*.3-sin(\x*50)*.4-4.5}) -- plot[domain=10.6:0] (\x, {-6}) -- cycle; 
\draw[-,color=BlueViolet,line width= .8mm] (0,-6) -- (10.6,-6);
\draw [color=Blue, smooth, line width=0.3mm, samples=100, domain=0:10.6] plot(\x,{cos(\x*80)*.3-sin(\x*50)*.4-4.5});
\draw[-to,color=Black,line width=0.3mm] (11,-5.8) -- (11,-4.2);
\node at (11.7,-5){$\rho(x)$};
\node at (11.5,-2.5){$n_i$};
\end{tikzpicture}
    \caption{Schematic of a one dimensional, periodic lattice gas and its hydrodynamic description. The top line represents a single configuration of particles, with the arrows denoting some possible hops. In the second line we have converted to occupation numbers, $n_i$. In the final line we show the smooth density that arises through coarse graining the occupation numbers. }
    \label{fig:Schematic}
\end{figure}

As sketched in \cref{fig:Schematic}, the state of a stochastic lattice gas is given by a configuration of particles, $\mathcal{C}$, sitting on the discrete sites of a lattice. Dynamics are introduced by letting particles move around or chemically react, changing the configuration to $\mathcal{C}'$ at rate $\mathcal{W}(\mathcal{C}\rar\mathcal{C}')$ \cite{cardy2006reaction,kipnis2013scaling,spohn2012large,landau2021guide}. 
This evolution is stochastic, and using these rates, the probability of seeing a given configuration, $\prob(\mathcal{C})$, follows the master equation \cite{van1992stochastic,gardiner1985handbook}
\begin{equation}
    \pd_t\prob(\mathcal{C})=\sum_{\mathcal{C}'}\mathcal{W}(\mathcal{C}'\rar\mathcal{C})\prob(\mathcal{C}')-\mathcal{W}(\mathcal{C}\rar\mathcal{C}')\prob(\mathcal{C}),
\label{eqn:master equation}
\end{equation} 

and, for example, the master equation describing a box of $n$ particles which may react to produce $\Delta n$ more at rate $r(n)$ is
\begin{equation}
    \pd_t \prob(n) = r(n-\Delta n)\prob(n-\Delta n)-r(n)\prob(n).
\end{equation}
To generalise this master equation from one box, to a full lattice, we first expose the linearity of the master equation by re-writing the equation in operator form
\begin{equation}
    \pd_t\prob(n) = \ham\prob,\quad \ham =\left(e^{-\Delta n\pd_n}-1\right)r(n),
\end{equation}
where we used the identity $e^{\pd_n}f(n)=f(n+1)$. The evolution operator, or Hamiltonian, $\ham$, is a product of two terms: the rate $r(n)$, and an exponential minus one, which controls the particle production. This structure is generic, and can be extended to act on a lattice gas by defining an occupation number $n_i$ for each lattice site $i$:
\begin{equation}
    \ham(\pd_{{n}_i},n_i) = \sum_i \left(e^{-\Sigma_i \Delta n_{i}\pd_{{n}_i}}-1\right)r_i(\{n\}),
\label{eqn: general hamiltonian}
\end{equation}
where the rate $r_i(\{n\})$ may depend on the occupations of all sites. Multiple species of particle can also be accommodated through an additional occupation number $n^{\alpha}$ for each species $\alpha$.

For complicated lattice gas dynamics, these Hamiltonians are much easier to write down than the full master equation. As a demonstration, let us consider the one dimensional lattice gas shown in \cref{fig:Schematic}, where the only process is diffusion. We model this as particles hopping symmetrically at rate $1$ onto neighbouring sites, with no limit on the number of particles per site. Using the general Hamiltonian of \cref{eqn: general hamiltonian} we write the Hamiltonian as the sum of two hopping terms, one which moves a particle to the neighbouring right site, and the other which takes a particle to the left
\begin{equation}
\ham =\sum_i \left(e^{-\pd_{n_{i+1}}+\pd_{n_{i}}}-1\right)n_i +\left(e^{-\pd_{n_{i-1}}+\pd_{n_{i}}}-1\right)n_i.
\label{eqn:diffusion discrete action}
\end{equation} 
Although this example was simple, the jump rates $\mathcal{W}$ can be cooked up to model almost any process, and we recommend Refs. \cite{cardy2006reaction,tauber2014critical,wiese2016coherent} for other interesting examples.

\subsection{Path integral}
With lattice gases introduced, we would now like to derive hydrodynamic equations and LDFs describing their macroscopic behaviour. Although not necessary, it is easiest to do this using the path integral because it naturally gives the trajectory probabilities. There are various path integral representations of master equations, perhaps most famous being that of Doi and Peliti \cite{doi1976second,peliti1985path,tauber2014critical,wiese2016coherent}, which works by converting the master equation to second quantised form and using coherent states. Not only is this algebraically cumbersome, it can be difficult to map Doi-Peliti variables to physical particle densities \cite{doipelitifail}. As such, we instead use a more direct method \cite{saha2023large,baek2015singularities}, similar to the approach introduced in Refs. \cite{andreanov2006field,lefevre2007dynamics}.

As in all path integrals, the key object of interest is the evolution operator taking the probability at time zero to the probability at time $T$. Similar to quantum mechanics, this is given by the exponential of the Hamiltonian, $\exp[\ham T]$, and as shown in \ref{sec:appendix path derivation}, this has the path integral representation
\begin{equation}
     \int\mathcal{D}[n_i,\nh_i] \exp[-\int^T_0 \wrt t\ \left(\sum_i\nh_i \pd_t n_i\right)-\ham(\nh_i,n_i)],
\label{eqn:discrete path integral}
\end{equation}
where $\hat{n}$ is the conjugate `momentum', taking the place of $-\pd_n$ in the Hamiltonian. In this representation, the general form of the Hamiltonian is even clearer, with
\begin{equation}
\ham(n_i,\hat{n}_i) = \sum_i \left(e^{\sum_i \Delta n_i \hat{n}_i}-1\right)r_i(\{n\}).
\label{eqn: general ham operator}
\end{equation}
This path integral fully captures the microscopic dynamics of the lattice gas, but to study the large scale, hydrodynamic behaviour, we must coarse grain the particle numbers $n_i$, into a smooth density field $\rho(x)$ \cite{bertini2015macroscopic,kipnis2013scaling}.

\subsection{Hydrodynamic limit}
\label{sec:hydro limit}

As the lattice spacing, $\ep$, tends to zero, we expect the discrete particle numbers $n_i$ to converge to a smooth hydrodynamic density profile $\rho(x=i\ep)$, which only varies on a macroscopic scale. To convert between these two levels of description, one may be tempted to take the discrete path integral (\ref{eqn:discrete path integral}), and directly replace $n_i$ with $\rho(x)$ everywhere. This `mean-field' approximation sometimes works, but, as shown in section four of Ref. \cite{solon2015flocking} and  Ref. \cite{doipelitifail}, generally gives incorrect hydrodynamic equations. The failure arises because this naive smoothing does not account for fluctuations/ correlations that may exist in small mesoscopic regions around each lattice site. Such correlations are captured by being more careful and considering the local equilibrium structure of the lattice gas \cite{kipnis2013scaling,spohn2012large,lefevre2007dynamics}.

To this end, we will first write the Hamiltonian as the sum of a Hamiltonian density, $\mathcal{H}=\sum_i H_i(\{n_i,\nh_i\})$, which is given by the term in the summation of (\ref{eqn: general ham operator}). We then divide the lattice up into mesoscopic boxes, of linear dimension $\delta$ centred on the lattice sites, where $\delta$ is much larger than the lattice spacing, $\ep$, but much smaller than the system size \cite{bertini2015macroscopic}. This lets us write the Hamiltonian as sum over boxes
\begin{equation}
    \mathcal{H}=\sum_{x}\frac{1}{|B_{\delta}(x)|}\sum_{i\in B_{\delta}(x)}H(\{\nh_i,n_i\}).
\end{equation}
where $|B_{\delta}(x)|$ is the number of sites in the box and must be included to avoid over counting. 

The assumption of local equilibrium then states that each of these boxes rapidly relaxes to a local equilibrium state that is near constant in each box, but which evolves slowly over the macroscopic, hydrodynamic scale \cite{bertini2015macroscopic,doyon2023ballistic,demasi1986reaction}. The average number of each particles in each box is then equal to the slowly varying density $\rho(x,t)$, and the empirical averages over each box are assumed to converge to the average over the (as yet unknown) local equilibrium measure. Mathematically this means
\begin{equation}
    \frac{1}{|B_{\delta}(x)|}\sum_{i\in B_{\delta}(x)}H(\{\nh_i,n_i\})\rar \langle H(\{\nh_i,n_i\})\rangle,
\end{equation}
where the angle brackets mean an average over local equilibrium. In general, we expect some form of local equilibrium to exist when the particles are so rapidly mixed in each box, that each particle can explore the entire mesoscopic box over a microscopically short time, while only leaving the box after a macroscopic time \cite{de1985rigorous}. This requirement holds for the  diffusively dominated dynamics studied in this work, and for the models we study the local equilibrium has been rigorously established \cite{kourbane2018exact,erignoux2021hydrodynamic}.

Assuming a slowly varying density profile, we exchange the sum over sites with an integral, yielding
\begin{equation}
        \ham(\rh,\rho) = \ep^{-d}\int\dr\ \langle H(\{\nh_i,n_i\})\rangle,
\end{equation}
where we have also relabelled $\nh$ to $\rh$, and used $d$ for the spatial dimension. Here we have also assumed that the conjugate field can be directly smoothed which we can only justify by comparing our results with other work \cite{agranov2023macroscopic}. In general we must also perform some extra steps like Taylor expanding the fields around each site, but this is better seen through example. 

Substituting these results into (\ref{eqn:discrete path integral}) yields the hydrodynamic path integral governing the stochastic evolution of the density field $\rho$
\begin{equation}
    \begin{split}
        &\mathcal{P}[\rho(\textbf{r},t)] = \int \mathcal{D}[\rho,\rh] e^{-\mathcal{A}/\ep^{d}},\quad
        \mathcal{A}=\int\wrt t\int\dr\ \rh\pd_t\rho-\ham(\rh,\rho).
    \end{split}
   \label{eqn: hydro path integral}
\end{equation}
The factor of $\ep^{-d}$ in the exponent means that, as $\ep$ tends to zero, the probability of observing a single trajectory takes large deviation form, $\prob\asymp \exp[-\mathcal{A}/\ep^d]$. We thus identify the action $\mathcal{A}$ as the LDF for a single trajectory. 

\subsection{Hydrodynamics for diffusion}
\label{sec: hydro diffusion}
Our discussion thus far has been rather formal, so we now demonstrate the approach using the one dimensional diffusion model introduced earlier. Using the conjugate variable, the Hamiltonian in \cref{eqn:diffusion discrete action} is re-written as
\begin{equation}
\ham =\sum_i \left(e^{\nh_{i+1}-\nh_i}+e^{\nh_{i-1}-\nh_i}-2\right)n_i.
\end{equation}
We then have to take the average over local equilibrium,. Because this Hamiltonian is linear in the occupation number, $n_i$, it turns out we do not need to know the exact equilibrium measure as $\langle n_i\rangle=\rho(x)$ by definition. However, as we need a similar result later, let us note that the local equilibrium measure is known to be product Poissonian \cite{bodineau2010current}, i.e.
\begin{equation}
\mu_{\text{eqm}}=\prod_i \frac{\rho_i^{n_i}e^{-\rho_i}}{n_i !},
\end{equation}
which follows quite naturally from the fact that we can have an arbitrary number of particles per site, and that they are non-interacting. Performing the average over $\ham$ we arrive at the same expression with $n$ swapped for $\rho$. Assuming the density is smooth we then exchange the sum for an integral and Taylor expand the fields to $O(\ep)$ about each site, yielding
\begin{equation}
    \ham=\ep\int\dr\ \rho\pd_x^2\rh +\rho(\pd_x\rh)^2+O(\ep^2).
\end{equation}
Substituting this Hamiltonian into the action, the final (optional) steps are to diffusively re-scale time according to $\wrt t\rar\ep^{-2}\wrt t$ and perform some integrations by parts. Doing this we arrive at the known LDF for non-interacting diffusing particles \cite{bertini2015macroscopic}
\begin{equation}
    \mathcal{A}=\int\dr\int\wrt t\ \rh(\pd_t\rho-\pd_x^2\rho)-\rho(\pd_x\rh)^2.
\end{equation}
Because this LDF is quadratic in $\hat{\rho}$, stochastic hydrodynamic equations can be extracted using the Martin--Siggia--Rose--Janssen--De Dominicis (MSRJD) formalism \cite{martin1973statistical,janssen1976lagrangean,dominicis1976techniques},  yielding
\begin{equation}
\pd_t\rho=\pd_x^2\rho+\pd_x\xi,
\end{equation}
where $\xi$ is a Gaussian noise with variance
\begin{equation}
\langle\xi(x,t)\xi(x',t')\rangle=2\ep\rho\delta(x-x')\delta(t-t').
\end{equation}
The scaling of the noise with the lattice spacing means that the deterministic hydrodynamic equations become exact in the limit of zero lattice spacing, and this will remain true for all the lattice gases we study. 

With the framework laid out we turn to the main focus of the text, active lattice gases.

\section{Motility induced phase separation}
\label{sec:MIPS}

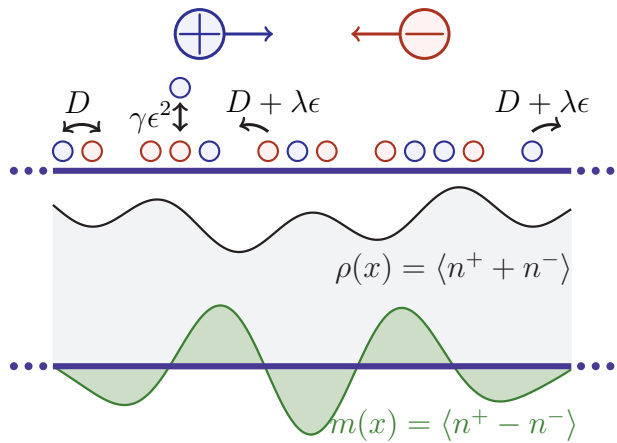
\begin{figure}[h]
    \centering
    \begin{tikzpicture}[scale=0.65]
\filldraw[color=Blue, fill=blue!5,line width=0.4mm] (3,2.8) circle (.5);
\node[scale=2,color=Blue,line width= .8mm] at (3,2.8){$+$};
\draw[-to,color=Blue,line width=0.4mm] (3.5,2.8) -- (4.5,2.8);
\filldraw[color=Mahogany, fill=red!5,line width=0.4mm] (7.6,2.8) circle (.5);
\node[scale=2,color=Mahogany] at (7.6,2.8){$-$};
\draw[-to,color=Mahogany,line width=0.4mm] (7.1,2.8) --(6.1,2.8);
\draw[-,color=BlueViolet,line width= .8mm] (0,0) -- (10.6,0);
\filldraw[color=BlueViolet, fill=BlueViolet,thick] (-0.2,0) circle (.06);
\filldraw[color=BlueViolet, fill=BlueViolet,thick] (-0.5,0) circle (.06);
\filldraw[color=BlueViolet, fill=BlueViolet,thick] (-0.8,0) circle (.06);
\filldraw[color=BlueViolet, fill=BlueViolet,thick] (10.8,0) circle (.06);
\filldraw[color=BlueViolet, fill=BlueViolet,thick] (11.1,0) circle (.06);
\filldraw[color=BlueViolet, fill=BlueViolet,thick] (11.4,0) circle (.06);
\filldraw[color=Blue, fill=blue!5,thick] (0.2,.4) circle (.2);
\filldraw[color=Mahogany, fill=red!5,thick] (0.8,.4) circle (.2);
\filldraw[color=Mahogany, fill=red!5,thick] (2,.4) circle (.2);
\filldraw[color=Mahogany, fill=red!5,thick] (2.6,.4) circle (.2);
\filldraw[color=Blue, fill=blue!5,thick] (3.2,.4) circle (.2);
\filldraw[color=Mahogany, fill=red!5,thick] (4.4,.4) circle (.2);
\filldraw[color=Blue, fill=blue!5,thick] (5,.4) circle (.2);
\filldraw[color=Mahogany, fill=red!5,thick] (5.6,.4) circle (.2);
\filldraw[color=Mahogany, fill=red!5,thick] (6.8,.4) circle (.2);
\filldraw[color=Blue, fill=blue!5,thick] (7.4,.4) circle (.2);
\filldraw[color=Blue, fill=blue!5,thick] (8,.4) circle (.2);
\filldraw[color=Mahogany, fill=red!5,thick] (8.6,.4) circle (.2);
\filldraw[color=Blue, fill=blue!5,thick] (9.8,.4) circle (.2);
\draw[-,color=BlueViolet,line width= .8mm] (0,-4) -- (10.6,-4);
\filldraw[color=BlueViolet, fill=BlueViolet,thick] (-0.2,-4) circle (.06);
\filldraw[color=BlueViolet, fill=BlueViolet,thick] (-0.5,-4) circle (.06);
\filldraw[color=BlueViolet, fill=BlueViolet,thick] (-0.8,-4) circle (.06);
\filldraw[color=BlueViolet, fill=BlueViolet,thick] (10.8,-4) circle (.06);
\filldraw[color=BlueViolet, fill=BlueViolet,thick] (11.1,-4) circle (.06);
\filldraw[color=BlueViolet, fill=BlueViolet,thick] (11.4,-4) circle (.06);
\draw [color=Black!5,fill=Black!5, smooth, samples=100, domain=0:10.6] plot(\x, {cos(\x*40)*.3-sin(\x*120)*.4-1}) -- plot[domain=10.6:0] (\x, {-4}) -- cycle; 
\draw [color=OliveGreen!5,fill=OliveGreen!20, smooth, samples=100, domain=0:10.6] plot(\x, {-sin(\x*85)+sin(\x*120)*.4-4}) -- plot[domain=10.6:0] (\x, {-4}) -- cycle; 
\draw [color=OliveGreen, smooth, line width=0.3mm, samples=100, domain=0:10.6] plot(\x,{-sin(\x*85)+sin(\x*120)*.4-4});
\draw [color=Black, smooth, line width=0.3mm, samples=100, domain=0:10.6] plot(\x,{{cos(\x*40)*.3-sin(\x*120)*.4-1}});
\draw[-,color=BlueViolet,line width= .8mm] (0,-4) -- (10.6,-4);
\node[color=Black] at (8.2,-2){$\rho(x)=\langle n^+ + n^-\rangle$};
\node[color=OliveGreen] at (8.2,-5.2){$m(x)=\langle n^+ - n^-\rangle$};
\draw[-to,color=Black, line width=.4mm] (9.8,.8) arc (130:70:.6);
\draw[-to,color=Black, line width=.4mm] (4.4,.8) arc (50:110:.6);
\draw[to-to,color=Black, line width=.4mm] (.2,.8) arc (130:50:.6);
\draw[to-to,color=Black, line width=.4mm] (2.6,.8) -- (2.6,1.4);
\filldraw[color=Blue, fill=blue!5,thick] (2.6,1.7) circle (.2);
\node at (.5,1.4){$D$};
\node at (2,1.1){$\gamma\ep^2$};
\node at (4.5,1.4){$D+\lambda\ep$};
\node at (10,1.4){$D+\lambda\ep$};
\end{tikzpicture}
    \caption{Schematic of the MIPS model. Blue particles are positive and propel right, red particles are negative and propel left, and there is only one particle allowed per site. The first line shows the possible microscopic moves. The second line shows the hydrodynamic fields, the grey region gives the total density $\rho$, and the green the magnetisation $m$.}
    \label{fig:MIPS schematic }
\end{figure}

In this section we study the hydrodynamic behaviour of a lattice model of MIPS first introduced in Ref. \cite{kourbane2018exact}. Similar to other run and tumble systems \cite{tailleur2008statistical}, it consists of two species of self-propelling particle $n^{\pm}$, each with a preferred direction of propulsion along the lattice. Hard-core interactions are then  ensured by only allowing one particle per site.

The microscopic dynamics include diffusion, where two neighbouring lattice sites swap occupants at rate $D$; self propulsion where plus (minus) particles hop right (left) onto empty sites at rate $\lambda\ep$; and `tumbling' where particles turn around by switching type at rate $\gamma\ep^2$. A schematic of these rules are shown in \cref{fig:MIPS schematic }. Simulations of this system show that at sufficiently high self propulsion and density, the system undergoes MIPS, and phase separates into a dense and dilute phase \cite{kourbane2018exact}.

To construct the hydrodynamic theory we start with the Hamiltonian, written down using the above rates and the general formula from \cref{eqn: general ham operator}. The result is
\begin{equation}
\begin{aligned}
\ham =&\sum_{|i-j|=\ep}\sum_{\alpha=\pm} D n_i^\alpha(1-n_j)\left(e^{\nh^\alpha_{j}-\nh^\alpha_{i}}-1 \right)+  Dn_i^{\alpha} n_j^{-\alpha}\left(e^{\nh^\alpha_{j}-\nh^\alpha_i -\nh^{-\alpha}_{j}+\nh^{-\alpha}_i}-1\right)\\
&+\lambda \ep n_i^{\alpha}(1-n_{i+\alpha})\left(e^{\nh^\alpha_{i+\alpha}-\nh^\alpha_{i}}-1\right)+\gamma\ep^2 n_i^\alpha\left(e^{\nh^{-\alpha}_i-\nh_i^\alpha}-1\right),
\end{aligned}
\label{eqn: discrete mips action}
\end{equation}
where we have defined the total number of particles on a site as $n_i=n^+_i+n^-_i$. The terms proportional to $D$ govern symmetric diffusion, to $\lambda$ govern self propulsion, and to $\gamma$ swap the species of each particle. Here we have written the Hamiltonian as a sum over the two species of particle $\alpha=\pm$, although when first constructing it, we found it easier to explicitly write down the terms for plus and minus particles separately, before simplifying it as a sum.

The Hamiltonian (\ref{eqn: discrete mips action}) describes the microscopic dynamics of the lattice gas, but to find understand its macroscopic behaviour we follow the steps of section \ref{sec:hydro limit}, and coarse grain. To do this this we need to know the local equilibrium measure, and it was found in Ref. \cite{kourbane2018exact}, using the observation that the scaling of the propulsion and tumble rates with $\ep$ means that they happen much less frequently, on a microscopic time scale, than the diffusive moves with rate $D$. As such, diffusion dominates microscopically and controls the local equilibrium \cite{bodineau2010current}. Knowing this, we use that the diffusion is equivalent to two coupled simple symmetric exclusion processes \cite{bodineau2010current,derrida2011microscopic}, as to a plus (minus) particle, the minus (plus) particles act just like empty sites. As proved rigorously in Ref. \cite{kourbane2018exact,erignoux2021hydrodynamic}, this implies that the local equilibrium measure is a product of Bernoulli distributions over each site for each species.
\begin{equation}
\mu_{\text{eqm}}=\prod_{i,\alpha}{(\rho_i^{\alpha})}^{n_i^{\alpha}}{(1-\rho_i^{\alpha})}^{(1-n^{\alpha}_i)}.
\end{equation} 

With the equilibrium measures established we now follow the remaining steps of section \ref{sec:hydro limit} to derive the hydrodynamic LDF. The algebra turns out to be extremely tedious and so we leave the full details to \ref{sec:appendix mips derivation}. To summarise the process: we first average over the equilibrium measures, then Taylor expand to leading order in $\ep$, before diffusively rescaling time: $\wrt t\rar \ep^{-2}\wrt t$. The result is simplified by changing to density and magnetisation variables: $\rho=\rho^{+}+\rho^{-}$, $m=\rho^{+}-\rho^{-}$, $\rh^+=(\rh+\hat{m})$, $\rh^-=(\rh-\hat{m})$. The density $\rho$ gives the average number of particles per site, whereas the magnetisation, $m$, measures the average alignment of all the particles. Substitution of these definitions yields the LDF
\begin{equation}
    \begin{aligned}
        \mathcal{A}=&\int\dr \int\wrt t\ \rh[\pd_t\rho+\lambda\pd_x(m(1-\rho))-D\pd_x^2\rho]
+\hat{m}[\pd_t m +\lambda\pd_x(\rho(1-\rho))-D\pd_x^2 m ]\\
& -D\rho(1-\rho)(\pd_x\rh)^2-D(\rho-m^2)(\pd_x\hat{m})^2
        -2Dm(1-\rho)(\pd_x\rh)(\pd_x\hat{m})\\
        &+\gamma\rho[1-\cosh 2\hat{m}]+\gamma m\sinh2\hat{m}.
    \end{aligned}
\label{eqn:MIPS LDF}
\end{equation}
This is precisely the known LDF that was derived in Refs. \cite{agranov2022entropy,agranov2023macroscopic} by mapping the dynamics to the ABC model for which the LDF is already known \cite{clincy2003phase,bodineau2011phase}. This LDF can be used to study both the deterministic hydrodynamic description, valid when $\ep$ tends to zero, and the fluctuations about these deterministic equations.

\subsection{Deterministic behaviour}
\label{sec:deterministic behaviour}
Before analysing fluctuations, it is instructive to look at the deterministic behaviour. Because this LDF is not Gaussian, we cannot directly use the MSRJD formalism of section \ref{sec: hydro diffusion} to extract the deterministic equations. Instead, we note that in the small lattice spacing limit, the path integral may be evaluated using the saddle point method \cite{bertini2015macroscopic}. Extremising the LDF, the relevant saddle point trajectories solve Hamilton's equations
\begin{equation}
    \begin{split}
        \pd_t\rho=\frac{\delta\ham}{\delta \rh}, \quad \pd_t \hat{\rho}=-\frac{\delta \ham}{\delta \rho},\\
        \pd_t m=\frac{\delta\ham}{\delta \hat{m}}, \quad \pd_t \hat{m}=-\frac{\delta \ham}{\delta m},
    \end{split}
\label{eqn: hamiltons equations}
\end{equation}
where the conjugate variables act like momenta \cite{bertini2015macroscopic}. Although complicated, to find the deterministic hydrodynamic equations we only need the trivial solution $\hat{\rho}=\hat{m}=0$, corresponding to noiseless, deterministic motion \cite{tauber2014critical,elgart2004rare,elgart2006classification,kamenev2023field,cardy2006reaction}. 

Substituting these into Hamilton's equations we arrive at a set of coupled PDEs for the fields $\rho,m$:
\begin{equation}
    \begin{gathered}
        \pd_t \rho +\lambda\pd_x m(1-\rho)=D\pd_x^2\rho,\\
        \pd_t m +\lambda\pd_x \rho(1-\rho) =D\pd_x^2 m -2\gamma m,
    \end{gathered}
\end{equation}
which are the hydrodynamic equations previously derived by more conventional methods in Ref. \cite{kourbane2018exact}.  In that work, simulations of these equations were also compared to those of the underlying lattice gas, and excellent agreement was found when the lattice spacing was on the order of $10^{-3}$.

\begin{figure}[t]
    \centering
    \includegraphics[width=.6\linewidth]{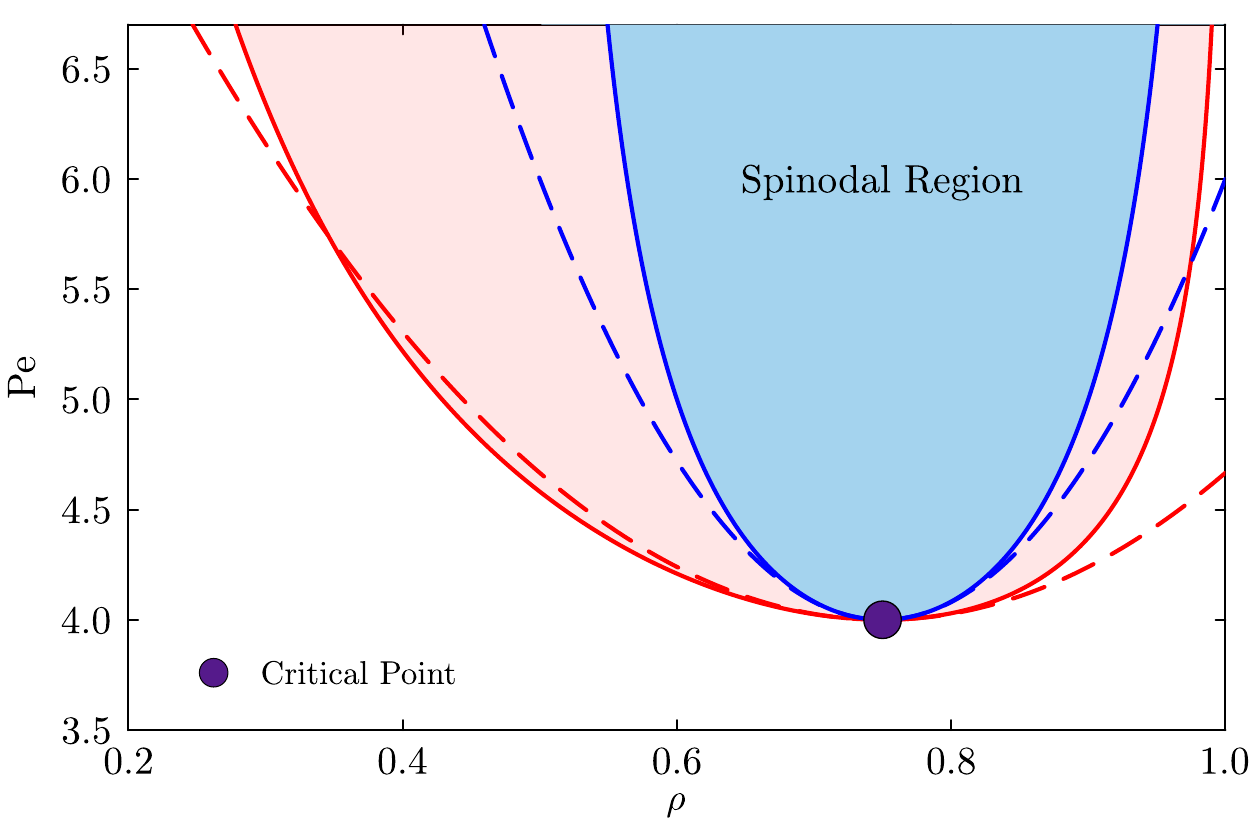}
    \caption{Phase diagram for MIPS. The solid blue line indicates the spinodal calculated from the hydrodynamic equations (\ref{eqn:deterministic mips 1}), above which homogeneous states are unstable and phase separate \cite{chaikin1995principles}. The red line is the binodal marking the phase separated densities, calculated in Ref. \cite{kourbane2018exact}. The dashed blue and red lines mark the spinodal and binodal curves from the $\phi^4$ free energy (\ref{eqn:mips phi4}), which agree with the exact curves near the critical point marked in purple.}
    \label{fig:mips critical}
\end{figure}

Following these authors, we simplify these equations by re-scaling space-time as $x\rar \sqrt{D/\gamma} x$, $t\rar t/\gamma$, arriving at
\begin{equation}
    \begin{gathered}
        \pd_t\rho+\text{Pe}\pd_x(m(1-\rho))=\pd_x^2\rho,\\
        \pd_t m+\text{Pe}\pd_x(\rho(1-\rho))=\pd_x^2m -2 m, 
    \end{gathered}
    \label{eqn:deterministic mips 1}
\end{equation}
where the P\'eclet number, comparing the rate of propulsion and diffusion, is $\text{Pe}=\lambda/\sqrt{D\gamma}$. Using these re-scaled equations, Kourbane-Housenne et al. \cite{kourbane2018exact} also calculated the spinodal and binodal densities shown in in \cref{fig:mips critical}, finding them to be in quantitative agreement with simulations of the underlying lattice gas.


\subsection{Connection to deterministic active model B}

With the deterministic hydrodynamics in hand, we now attempt to connect them to a more common theory of MIPS, active model B (AMB) \cite{tjhung2018cluster}.

Active model B is a field theory for MIPS written in terms of a single variable $\phi$, measuring the difference to the critical density \cite{cates2019active}. Although originally introduced phenomenologically, derivations from other microscopic models have shown that its predictions are most accurate near the critical point \cite{solon2018generalized}. Inspired by this, we apply a weakly non-linear analysis near the critical point to see if the lattice gas hydrodynamics can also be mapped onto AMB \cite{drazin1992nonlinear}. For those unfamiliar with weakly non-linear theory, it may be thought of as a formal approach to the standard Landau-Ginzburg expansion that is ordinarily performed near a critical point \cite{kardar2007statisticalfields,tauber2014critical}.

Before embarking on this calculation let us briefly redo the linear stability analysis that leads to the spinodal curve and critical point in \cref{fig:mips critical} \cite{kourbane2018exact}. As the spinodal marks the densities where uniform states are unstable, we linearise about uniform un-magnetised states with $\rho=\rho_0$ and $m=0$. Using fourier modes, we let
\begin{equation}
\rho=\rho_0 +\delta \rho e^{ikx+\sigma t},\quad m = \delta m e^{ikx+\sigma t},
\end{equation}
where $k$ is the wave-number of the perturbation, and $\sigma$ is the growth rate. Substituting these into (\ref{eqn:deterministic mips 1}) and keeping terms that are leading order in $\delta$, we find the growth rates
\begin{equation}
\sigma_{\pm} = -1-k^2\pm \sqrt{1-k^2 \text{Pe}^2\left(1-3 \rho+2 \rho^2\right)},
\label{eqn:mips growth rate}
\end{equation}
which are always negative below the critical P\'eclet number of $4$. As the P\'eclet number is increased past this critical value, the first density to become unstable is $\rho_0=3/4$, and beyond this, all densities inside the blue spinodal region of \cref{fig:mips critical} are unstable. Thus, the position of the critical point is $(\text{Pe}_c,\rho_c)=(4,3/4)$. Because these equations are deterministic, it is perhaps best to refer to this point not as a critical point, but rather as a bifurcation point associated with the hydrodynamic equations (\ref{eqn:deterministic mips 1}). However, this lattice gas, and all those studied in this work, is constructed such that the deterministic description becomes asymptotically exact in the limit of zero lattice spacing, which here is the appropriate thermodynamic limit relevant to the phase transition \cite{clincy2003phase,bertini2015macroscopic}, and so we believe the critical point terminology to be valid.

\begin{figure}[t]
    \centering
    \begin{tikzpicture}[scale=2.5,domain=0:2.05]
\draw[-stealth,color=gray,line width = .7mm] (-.5,0) -- (2.3,0); \node[left,black] at (2.5,0) {$k$};
\node[left,black] at (0,1.8) {$\sigma$};
\draw[-stealth,color=gray,line width = .7mm] (0,0) -- (0,1.8);
\draw[Blue,loosely dashed, line width = .4mm] (1.41,0) node[anchor=north]{$k= O(r^{1/2})$}--(1.41,1.5);
\node[below,Blue] at (-.5,1.5) {$\sigma_{\text{max}}=O(r^2)$};
\draw[blue,loosely dashed, line width = .4mm] (0,1.5)--(1.41,1.5);
\draw[ line width =.7mm,black, smooth]   plot (\x,{6*(\x/2)^2-6*(\x/2)^4});
\end{tikzpicture}
     \caption{Positive growth rate, $\sigma_+$, against wave-number near the critical point, $\text{Pe}=4,\rho=3/4$. The growth rate tends to zero at $k=0$, has a peak that scales like $r^2$, and the peak wave-number scales like $r^{1/2}$.}
    \label{fig: Growth rate}
\end{figure}
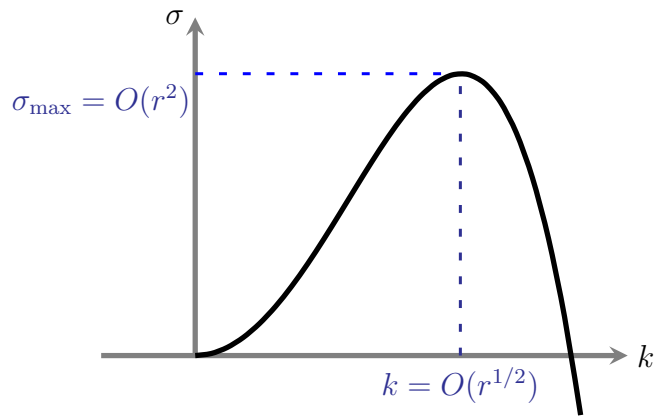

To examine the behaviour near the critical point, we set $\text{Pe}=4(1+r)$, where $r$ is a small parameter, analogous to the reduced temperature in equilibrium \cite{kardar2007statisticalfields}. Expanding the growth rate (\ref{eqn:mips growth rate}) in powers of $k$ and $r$, we find
\begin{equation}
\sigma_{+}\sim 2k^2 r -\frac{1}{2}k^4 + O(k^2 r^2),
\end{equation}
which, as shown in \cref{fig: Growth rate}, tells us that the dominant wave-number scales like $r^{1/2}$, while the peak scales like $r^2$. As is standard in weakly non-linear theory \cite{cross1993pattern,drazin1992nonlinear,drazin2002introduction}, these scalings inform our choice of slow space-time variables, and so we set $X=r^{1/2} x$, and $\tau=r^2 t$.

With these slow variables, we asymptotically expand the density and magnetisation fields in $r$
\begin{equation}
    \begin{gathered}
        \rho=\frac{3}{4}+r^{1/2}\rho_1(X,\tau) + r\rho_2(X,\tau) + ...,\\
        m = r^{1/2}m_1(X,\tau) + r m_2(X,\tau) + ....,
    \end{gathered}
\label{eqn: MIPS asymptotic expansion}
\end{equation}
and substitute them into the deterministic equations (\ref{eqn:deterministic mips 1}), before solving them order by order in $r$. From the magnetisation equation of (\ref{eqn:deterministic mips 1}), we find that the magnetisation is slaved to the density, with
\begin{gather}
    m_1=0,\quad m_2=\pd_X\rho_1,\quad m_3 =\pd_X(\rho_2+2\rho_1^2),\\
    m_4 = \pd_X(\rho_1+\rho_3)+4\pd_X(\rho_1\rho_2)+\frac{1}{2}\pd_X^3\rho_1
\label{eqn:slaved mag}
\end{gather}
found from solving the equations up to $O(r^2)$. Substituting these into the density equation from  (\ref{eqn:deterministic mips 1}), we find the first non-zero term at $O(r^{5/2})$ to be
\begin{equation}
    \pd_\tau \rho_1 = 32\rho_1(\pd_X\rho_1)^2 -2\pd_X^2\rho_1 +16\rho_1^2\pd_X^2\rho_1-\frac{1}{2}\pd_X^4\rho_1.
\label{eqn:slow expansion equation}
\end{equation}
Converting back to the original space-time variables and letting $\phi=\rho
-\rho_c$, we can put \cref{eqn:slow expansion equation} in equilibrium model B form \cite{hohenberg1977theory}
\begin{equation}
    \pd_t\phi=\pd_x\left(\pd_x\frac{\delta \mathcal{F}}{\delta \phi}\right),
\label{eqn: mips model b}
\end{equation}
where the free energy $\mathcal{F}$ takes $\phi^4$ form
\begin{equation}
    \mathcal{F}[\phi]=\int\dr\ -r\phi^2 +\frac{4}{3}\phi^4+\frac{1}{4}(\pd_x\phi)^2.
\label{eqn:mips phi4}
\end{equation}
These equations capture the dynamics near the critical point, and as shown in \cref{fig:mips critical}, the spinodal and binodal curves from this free energy agree with the full curves calculated from \cref{eqn:deterministic mips 1} near the critical point. We have therefore shown that the lattice gas lies in the mean-field model B universality class when the lattice spacing tends to zero. Interestingly, this could have been spotted already from the slow space time variables, as a growth rate scaling like $r^z$ for $z=2$, and a wave-number scaling like $r^{1/2}$ indicates a mean-field model B class \cite{hohenberg1977theory}.

Of course, model B is not equivalent to AMB, however it can be derived from (\ref{eqn:deterministic mips 1}) by continuing the calculation to higher order in $r$ as this captures the dynamics further away from the critical point. At $O(r^{5/2})$, the magnetisation equation gives
\begin{equation}
m_5=\pd_X\left[2\rho_1^2 +2\rho_2^2 + \rho_2 +4\rho_1\rho_3 + \rho_4 +2\pd_X(\rho_1\pd_x\rho_1) +\frac{1}{2}\pd_X^2 \rho_2 \right],
\end{equation}
which upon substitution into the density equation from (\ref{eqn:deterministic mips 1}), yields
\begin{equation}
\pd_\tau\rho_2 = \pd_X^2\left(2\rho_2-16\rho_1^2\rho_2 -3(\pd_X\rho_1)^2-\frac{1}{2}\pd_X^2\rho_2 \right).
\end{equation}
Multiplying this equation by $r^3$ and adding it to equation (\ref{eqn:slow expansion equation}) multiplied by $r^{5/2}$, we find that certain terms can be re-summed using the asymptotic expansion \ref{eqn: MIPS asymptotic expansion}. The result is
\begin{equation}
    \pd_t\phi = \pd_x^2\left(\frac{\delta \mathcal{F}}{\delta \phi}-3(\pd_x\phi)^2\right),
\label{eqn:ambp mips}
\end{equation}
which is exactly AMB, with the $\lambda$ parameter controlling the non-equilibrium term equal to $-3$ \cite{wittkowski2014scalar,tjhung2018cluster}. This is satisfying as active model B is normally derived non-rigorously, whereas here it \textit{exactly} captures the dynamics just away from the critical point.

\subsection{Near critical probability distribution}

Having seen how the deterministic equations compare to (active) model B, we now analyse the effect of fluctuations, which are important for all finite, but non-zero lattice spacings. We will limit ourselves to extremely small lattice spacings where, as the noise scales like $\ep$ (c.f. section \ref{sec: hydro diffusion}), the noise is weak. Importantly, the noise remains weak even near the critical point \cite{clincy2003phase}, which is quite different from the situation considered in ordinary statistical mechanics, where fluctuations are magnified near the critical point \cite{chaikin1995principles,goldenfeld2018lectures,kardar2007statisticalfields}.

Our approach will use the techniques of weak-noise theory \cite{bray1990path,graham1985weak}, with the aim to perturbatively calculate the stationary probability distribution near the critical point. In weak-noise theory the probability of observing a given configuration is given by the probability of the most likely state, multiplied by the transition probability of moving between the most-likely state and the desired configuration in an infinite amount of time \cite{bray1990path,bertini2015macroscopic,baek2015singularities},i.e.
\begin{equation}
\prob[\config] \asymp \prob[\config_0]\prob[\config_0\rar \config],
\label{eqn:weak noise stationary formula}
\end{equation}
where $\config_0$ is the most likely configuration.

This transition probability is readily computed using the path integral (\ref{eqn: hydro path integral}) by conditioning the paths on the desired initial and final configurations. As mentioned in section \ref{sec:deterministic behaviour}, the path integral is dominated by the saddle point trajectories which solve Hamilton's equations (\ref{eqn: hamiltons equations}). However, unlike the deterministic case, we will be interested in fluctuations where $\hat{\rho},\hat{m}$ may not be zero \cite{elgart2004rare}. The initial condition for these equations is the most likely configuration which must be a fixed point of Hamilton's equations (\ref{eqn: hamiltons equations}). 

The general form of any LDF, and therefore Hamilton's equations are such that any fixed point must have $\hat{\rho},\hat{m}=0$, i.e. the fixed-point must be noise-less \cite{baek2015singularities}. Because the Hamiltonian is conserved by the dynamics, we find
\begin{equation}
    \prob[\config_0\rar \config] \asymp \exp[-\frac{\sqrt{D}}{\sqrt{\gamma}\ep}\int^{\infty}_0 \wrt t\int\dr\ \rh\pd_t\rho+\hat{m}\pd_t m],
\label{eqn: zero ham transition}
\end{equation}
with $\hat{\rho},\hat{m}$ evaluated along the saddle path connecting configurations $\mathcal{C}_0$ and $\mathcal{C}$. The pre-factor $\sqrt{D/\gamma}$ comes from converting to the dimensionless units of \cref{eqn:deterministic mips 1}, which handily sets all other coefficients in the LDF to unity \cite{agranov2023macroscopic}. Unfortunately this simplification does not make solving (\ref{eqn: hamiltons equations}) any easier, and similar saddle equations have only been solved in a handful of cases (see Refs. \cite{tailleur2008mapping,derrida2011microscopic,saha2023large} for interesting examples).

\begin{figure}[t]
\centering
\begin{subfigure}[t]{0.02\textwidth}
	\text{(a)}
\end{subfigure}\hfill
\begin{subfigure}[t]{0.45\textwidth}
  \centering
  \begin{adjustbox}{valign=t}
  \begin{tikzpicture}[scale=0.45]
\draw[-stealth,color=gray,line width= .7mm] (0,-5) -- (0,5);
\draw[-stealth,color=gray,line width= .7mm] (-7,0) -- (7,0);
\node[scale=1.3] at (1,5){$\hat{\rho}$};
\node[scale=1.3] at (7,1){$\rho$};
\draw [color=Mahogany!5,fill=Mahogany!5, smooth, samples=100, domain=0:6] plot(\x, {-10*(\x/6)+10*(\x/6)^3}) -- plot[domain=6:0] (\x, {0}) -- cycle; 
\draw [color=Mahogany!5,fill=Mahogany!5, smooth, samples=100, domain=-6:0] plot(\x, {-10*(\x/6)+10*(\x/6)^3}) -- plot[domain=-6:0] (\x, {0}) -- cycle; 
\draw [-stealth, color=Mahogany, smooth, samples=100, domain=-6:-3,line width= .7mm] plot(\x,{-10*(\x/6)+10*(\x/6)^3});
\draw [color=Mahogany, smooth, samples=100, domain=-3:0,line width= .7mm] plot(\x,{-10*(\x/6)+10*(\x/6)^3});
\draw[-stealth, color=blue, line width=0.7mm] (0,0)--(-3,0);
\draw[color=blue, line width=0.7mm] (-2,0)--(-6,0);
\draw[-stealth, color=blue, line width=0.7mm] (0,0)--(3,0);
\draw[color=blue, line width=0.7mm] (2,0)--(6,0);
\draw [-stealth, color=Mahogany, smooth, samples=100, domain=6:3,line width= .7mm,] plot(\x,{-10*(\x/6)+10*(\x/6)^3});
\draw [color=Mahogany, smooth, samples=100, domain=3:0,line width= .7mm] plot(\x,{-10*(\x/6)+10*(\x/6)^3});
\filldraw[color=Black, fill=Black,thick] (0,0) circle (.2);
\filldraw[color=Black, fill=Black,thick] (-6,0) circle (.2);
\filldraw[color=Black, fill=Black,thick] (6,0) circle (.2);
\end{tikzpicture}
  \end{adjustbox}
\end{subfigure}\hfill
\begin{subfigure}[t]{0.02\textwidth}
	\text{(b)}
\end{subfigure}\hfill
\begin{subfigure}[t]{.45\textwidth}
  \centering
  \begin{adjustbox}{valign=t}
  \begin{tikzpicture}[scale=0.45,domain=-5.5:5.5]
\draw[-stealth,color=gray,line width= .7mm] (0,-5) -- (0,5);
\draw[-stealth,color=gray,line width= .7mm] (-7,0) -- (7,0);
\node[scale=1.3] at (1,5){$\mathcal{F}$};
\node[scale=1.3] at (7,1){$\rho$};
\draw[ line width =.7mm,black, smooth]   plot (\x,{10*(-(\x/5)^2+(\x/5)^4))});
\draw[-, domain=2.0:3.536, smooth, blue,line width =.8mm]  plot (\x,{10*(-(\x/5)^2+(\x/5)^4))});
\draw[-stealth, domain=0:2.0, smooth, blue,line width =.8mm]  plot (\x,{10*(-(\x/5)^2+(\x/5)^4))});
\draw[-stealth, domain=-3.536:-2.0, smooth, variable=\x, Mahogany,line width =.8mm]  plot (\x,{10*(-(\x/5)^2+(\x/5)^4))});
\draw[-, domain=-2.0:0.0, smooth, variable=\x, Mahogany,line width =.8mm]  plot (\x,{10*(-(\x/5)^2+(\x/5)^4))});
\filldraw[color=Black, fill=Black,thick] (0,0) circle (.2);
\filldraw[color=Black, fill=Black,thick] (-3.536,-2.5) circle (.2);
\filldraw[color=Black, fill=Black,thick] (3.536,-2.5) circle (.2);
\end{tikzpicture}
\end{adjustbox}
\end{subfigure}
\caption{(a) Schematic zero Hamiltonian trajectories just above the critical point with $r>0$. The thick black dots mark the fixed points, which are all noiseless with $\hat{\rho}=0$. The central point is the homogeneous state with $\rho=\rho_c$, and the points either side are the phase separated densities. The fluctuating trajectories that increase the free energy are red, and have $\hat{\rho}\neq0$. The blue lines that decrease the free energy are deterministic, with $\hat{\rho}=0$. The area of the shaded region gives the action required to transition between each fixed point, which for a system in equilibrium is equivalent to the free energy difference. (b) Schematic of how the trajectories affect the free energy, with the three black dots corresponding, in order, to the black dots in (a).}
\label{fig: free energy trajectories}
\end{figure}
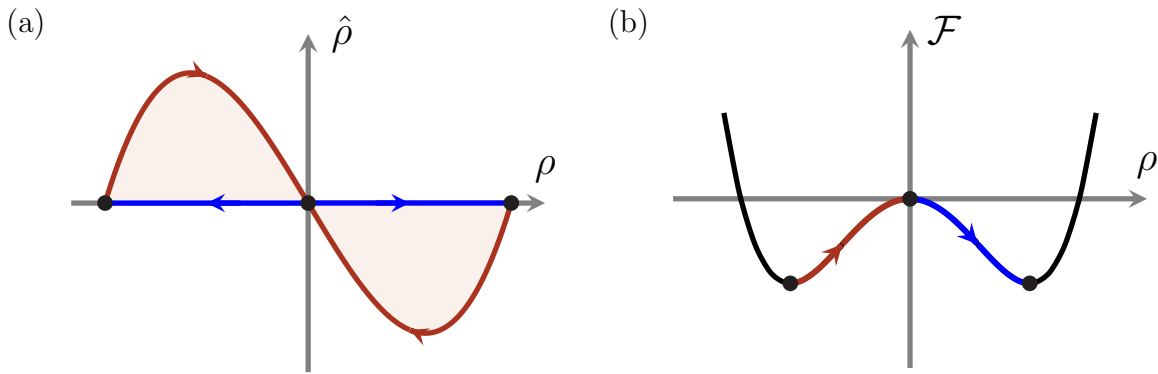

In our case we are saved in that we only wish to solve (\ref{eqn: hamiltons equations}) perturbatively in the small parameter $r$ introduced earlier. As in the deterministic case, we first convert to the slow space-time variables $X,\tau$. We then expand $\rho,m$ as in (\ref{eqn: MIPS asymptotic expansion}), while also expanding the conjugate variables
\begin{equation}
    \begin{gathered}
        \hat\rho=r^{1/2}\hat \rho_1(X,\tau) + r\hat\rho_2(X,\tau) + ...,\\
        \hat m = r^{1/2}\hat m_1(X,\tau) + r \hat m_2(X,\tau) + ....
    \end{gathered}
\label{eqn:conjugate expansion}
\end{equation}
Substituting these into (\ref{eqn: hamiltons equations}) and solving order by order, with the $\ham=0$ constraint, we find the conjugate variable are zero up to $O(r)$, with the first non-zero term
\begin{equation}
\begin{gathered}
    \hat{\rho}_3=\frac{16}{3}\left(-2\rho_1 + \frac{16}{3}\rho_1^3 -\frac{1}{2}\pd_X^2\rho_1 \right),\\
    \hat{m}_3 = 0.
\end{gathered}
\end{equation}
Converting back to the original space-time variables and again using $\phi=\rho-\rho_c$, we have
\begin{equation}
\hat{\rho} = \frac{16}{3}\frac{\delta \mathcal{F}}{\delta\phi}+O(r^2),
\end{equation}
which upon substitution into (\ref{eqn: zero ham transition}) gives
\begin{equation}
    \prob[\config_0\rar\config] \asymp \exp[ -\frac{16\sqrt{D}}{3\ep\sqrt{\gamma}} (\mathcal{F}[\config]-\mathcal{F}[\config_0] ) +O(r^2)].
\label{eqn:free energy fluctuation}
\end{equation}
Importantly, this calculation only holds if the final state has a larger free energy that the initial state; and as shown in \cref{fig: free energy trajectories}, this is because the fluctuating trajectories that leave the fixed points always serve to increase the free energy. In contrast, the trajectories that decrease the free energy are always deterministic with $\hat{\rho}=\hat{m}=0$ \cite{tailleur2007mapping,tailleur2008mapping}. Substituting these results into (\ref{eqn:weak noise stationary formula}) and re-arranging, we find that, up to an overall pre-factor that we cannot compute \cite{bray1990path}, the probability distribution for $\phi$ is equilibrium-like near the critical point, i.e.
\begin{equation}
    \prob[\config]\asymp \exp\left[ -\frac{16 \sqrt{D}}{3\sqrt{\gamma}\ep} \mathcal{F}[\phi(x)] +O(r^2) \right].
\label{eqn: mips boltzmann}
\end{equation}
This calculation confirms that the critical dynamics of this model is in the mean-field Ising universality class. Let us also note that the quadratic approximation to this free energy was calculated in Ref. \cite{agranov2021exact} using a Gaussian approximation to the full LDF.

Unfortunately we have not found it possible to calculate any of the corrections to this result that appear at higher order in $r$, as non-integrable AMB type terms start to arise \cite{cates2023classical}. \cref{eqn:ambp mips,eqn: mips boltzmann} are the key results of the paper, and in the next sections we perform similar calculations for two other models.

\section{Quorum sensing MIPS}
\label{sec:quorum}
\begin{figure}
    \centering
    \begin{tikzpicture}[scale=0.65]
\draw[-,color=BlueViolet,line width= .8mm] (0,0) -- (10.6,0);
\filldraw[color=BlueViolet, fill=BlueViolet,thick] (-0.2,0) circle (.06);
\filldraw[color=BlueViolet, fill=BlueViolet,thick] (-0.5,0) circle (.06);
\filldraw[color=BlueViolet, fill=BlueViolet,thick] (-0.8,0) circle (.06);
\filldraw[color=BlueViolet, fill=BlueViolet,thick] (10.8,0) circle (.06);
\filldraw[color=BlueViolet, fill=BlueViolet,thick] (11.1,0) circle (.06);
\filldraw[color=BlueViolet, fill=BlueViolet,thick] (11.4,0) circle (.06);
\filldraw[color=Blue, fill=blue!5,thick] (0.2,.4) circle (.2);

\filldraw[color=Blue, fill=blue!5,thick] (0.8,.4) circle (.2);
\filldraw[color=Blue, fill=blue!5,thick] (0.8,1) circle (.2);
\filldraw[color=Blue, fill=blue!5,thick] (1.4,.4) circle (.2);
\filldraw[color=Mahogany, fill=red!5,thick] (2,.4) circle (.2);
\filldraw[color=Mahogany, fill=red!5,thick] (2,1) circle (.2);
\filldraw[color=Blue, fill=blue!5,thick] (1.4,1) circle (.2);
\filldraw[color=Mahogany, fill=red!5,thick] 
(2.6,.4) circle (.2);
\filldraw[color=Blue, fill=blue!5,thick] 
(2.6,1) circle (.2);
\filldraw[color=Mahogany, fill=Red!5,thick] (3.2,.4) circle (.2);
\filldraw[color=Blue, fill=blue!5,thick] 
(3.2,1) circle (.2);
\filldraw[color=Blue, fill=blue!5,thick] 
(4.4,.4) circle (.2);
\filldraw[color=Blue, fill=blue!5,thick] 
(4.4,1) circle (.2);
\filldraw[color=Blue, fill=blue!5,thick] 
(4.4,1.6) circle (.2);
\filldraw[color=Blue, fill=blue!5,thick] 
(3.8,.4) circle (.2);
\filldraw[color=Blue, fill=blue!5,thick] 
(3.8,1) circle (.2);
\filldraw[color=Blue, fill=blue!5,thick] 
(3.8,1.6) circle (.2);
\filldraw[color=Mahogany, fill=red!5,thick] (5,.4) circle (.2);
\filldraw[color=Mahogany, fill=red!5,thick] (5,1) circle (.2);
\filldraw[color=Mahogany, fill=red!5,thick] (5,1.6) circle (.2);
\filldraw[color=Mahogany, fill=red!5,thick] (5.6,.4) circle (.2);
\filldraw[color=Mahogany, fill=red!5,thick] (5.6,1) circle (.2);
\filldraw[color=Mahogany, fill=red!5,thick] (6.2,.4) circle (.2);
\filldraw[color=Mahogany, fill=red!5,thick] (6.2,1) circle (.2);
\filldraw[color=Mahogany, fill=red!5,thick] (6.8,.4) circle (.2);
\filldraw[color=Blue, fill=blue!5,thick] (7.4,.4) circle (.2);
\filldraw[color=Blue, fill=blue!5,thick] (8,.4) circle (.2);
\filldraw[color=Mahogany, fill=red!5,thick] (8.6,.4) circle (.2);
\filldraw[color=Mahogany, fill=red!5,thick] (9.8,.4) circle (.2);
\draw[-,color=BlueViolet,line width= .8mm] (0,-4) -- (10.6,-4);
\filldraw[color=BlueViolet, fill=BlueViolet,thick] (-0.2,-4) circle (.06);
\filldraw[color=BlueViolet, fill=BlueViolet,thick] (-0.5,-4) circle (.06);
\filldraw[color=BlueViolet, fill=BlueViolet,thick] (-0.8,-4) circle (.06);
\filldraw[color=BlueViolet, fill=BlueViolet,thick] (10.8,-4) circle (.06);
\filldraw[color=BlueViolet, fill=BlueViolet,thick] (11.1,-4) circle (.06);
\filldraw[color=BlueViolet, fill=BlueViolet,thick] (11.4,-4) circle (.06);
\draw [color=black!5,fill=black!5, smooth, samples=100, domain=0:10.6] plot(\x, {sin(\x*25)*.3-2}) -- plot[domain=10.6:0] (\x, {-4}) -- cycle; 
\draw [color=OliveGreen!5,fill=Green!20, smooth, samples=100, domain=0:10.6] plot(\x, {cos(\x*15)*2-4}) -- plot[domain=10.6:0] (\x, {-4}) -- cycle; 
\draw [color=OliveGreen, smooth, line width=0.3mm, samples=100, domain=0:10.6] plot(\x,{cos(\x*15)*2-4});
\draw [color=black, smooth, line width=0.3mm, samples=100, domain=0:10.6] plot(\x,{{sin(\x*25)*.3-2}});
\draw[-,color=BlueViolet,line width= .8mm] (0,-4) -- (10.6,-4);
\node[color=black] at (8.5,-3){$\rho(x)=\langle n^+ + n^-\rangle$};
\node[color=OliveGreen] at (2.3,-5){$m(x)=\langle n^+ - n^-\rangle$};
\draw[-to,color=Black, line width=.4mm] (5,2) arc (50:110:.6);
\draw[-to,color=Black, line width=.4mm] (.8,1.4) arc (130:70:.6);
\draw[-to,color=Black, line width=.4mm] (9.8,.8) arc (130:70:.6);
\node at (.8,1.8){$D+e^{-2\sigma_0}\lambda\ep$};
\node at (5,2.5){$D+e^{-3\sigma_0}\lambda\ep$};
\node at (9.8,1.2){$D$};
\end{tikzpicture}
    \caption{Schematic of the quorum sensing MIPS model. Multiple particles are allowed per site, but the self propulsion is exponentially slower at regions of high density. Blue particles are positive, red particles are negative. The first line shows the microscopic moves. The second line shows the hydrodynamic fields, the grey region gives the total density, and the green the magnetisation.}
    \label{fig:MIPS Quorum }
\end{figure}
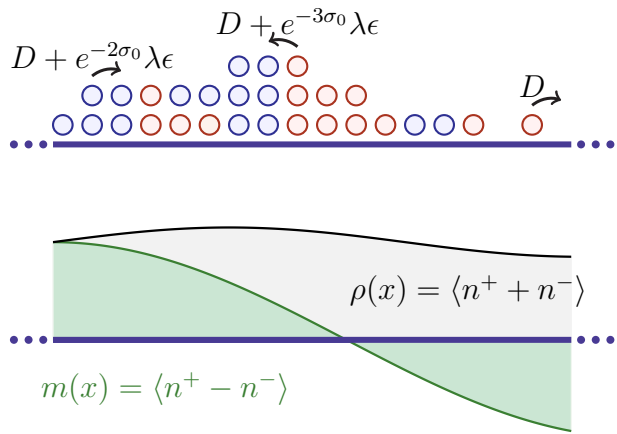

We now turn to an alternative model of MIPS based on Quorum sensing to argue that our conclusions about Ising universality are not model dependent. Quorum sensing is a way of modelling signalling between particles, with particles no-longer interacting in a hard-core way, but instead slowing down in regions of high density \cite{cates2015motility,tailleur2008statistical}. To put this model on a lattice we use a near identical set-up to that of section \ref{sec:MIPS} with two types of particle, but allow multiple particles per site. Each particle then diffuses independently along the lattice at rate $D$, but self propels at a rate $\ep \lambda_0 e^{-\sigma_0 (n^++n^-)}$ which is slower when the density is high (see \cref{fig:MIPS Quorum }). Here we have chosen the slowing down function, or `interaction kernel' as an exponential to ease calculation, however almost any short range kernel will suffice \cite{thompson2011lattice}. This self-propulsion speed depends only on the on-site density, i.e. the process is zero-range \cite{thompson2011lattice}, but interactions that include neighbouring particles can be included. However these act solely to scale $\sigma_0$ in the hydrodynamic limit and so can be safely ignored.

The microscopic Hamiltonian for this system is similar to (\ref{eqn: discrete mips action}), being
\begin{equation}
\begin{aligned}
\ham =&\sum_{|i-j|=\ep}\sum_{\alpha=\pm} D n_i^\alpha\left(e^{\nh^\alpha_{j}-\nh^\alpha_{i}}-1 \right)+\lambda \ep n_i^{\alpha}e^{-\sigma_0 n_i}\left(e^{\nh^\alpha_{i+\alpha}-\nh^\alpha_{i}}-1\right)+\gamma\ep^2 n_i^\alpha\left(e^{\nh^{-\alpha}_i-\nh_i^\alpha}-1\right),
\end{aligned}
\label{eqn:micro quorum hamiltonian}
\end{equation}
which is a sum of a diffusion, propulsion, and tumbling term. To coarse grain this Hamiltonian we need the local equilibrium measure, which, as in section \ref{sec:MIPS}, is controlled by the diffusion term alone. Because each species diffuses independently of the other, the local equilibrium measure is the same as that of the simple diffusion model in section (\ref{sec: hydro diffusion}), i.e. product poissonian with
\begin{equation}
\mu_{\text{eqm}}=\prod_{i,\alpha}\frac{{\rho^\alpha_i}^{n_i^\alpha} e^{-\rho_i^\alpha}}{n^\alpha_i !}.
\end{equation}
Averaging the microscopic Hamiltonian (\ref{eqn:micro quorum hamiltonian}) over this measure and performing a Taylor expansion, we arrive at the hydrodynamic LDF
\begin{equation}
    \begin{aligned}
        \mathcal{A}=&\int\dr\int\wrt t\ \rh[\pd_t\rho+\lambda\pd_x m e^{-\sigma\rho}- D\pd_x^2 \rho]+\hat{m}[\pd_t m +\lambda\pd_x \rho e^{-\sigma\rho}\\
        &-D\pd_x^2 m]-D\rho(\pd_x\rh)^2-D\rho(\pd_x\hat{m})^2-2Dm(\pd_x \hat{\rho})(\pd_x \hat{m})
        \\
        &+\gamma\rho[1-\cosh 2\hat{m}]+\gamma m\sinh2\hat{m},
    \end{aligned}
\label{eqn:quorum action}
\end{equation}
from which we see that the the average over local equilibrium has renormalised the self propulsion to $\lambda=\lambda_0e^{-\sigma}$, and $\sigma_0$ to $\sigma=1-e^{-\sigma_0}$. Interestingly, we found that this LDF could also be derived using the Doi-Peliti formalism if one performs the usual Cole-Hopf transform to change from coherent state variables to densities \cite{tauber2014critical,wiese2016coherent,scandolo2023active}. Such a connection is not expected to hold for all lattice gases, only happening here because the coherent states used in Doi-Peliti encode Poisson distributions \cite{cardy2006reaction,tauber2014critical}.

\subsection{Deterministic behaviour}
\begin{figure}[t]
    \centering
    \includegraphics[width=.5\linewidth]{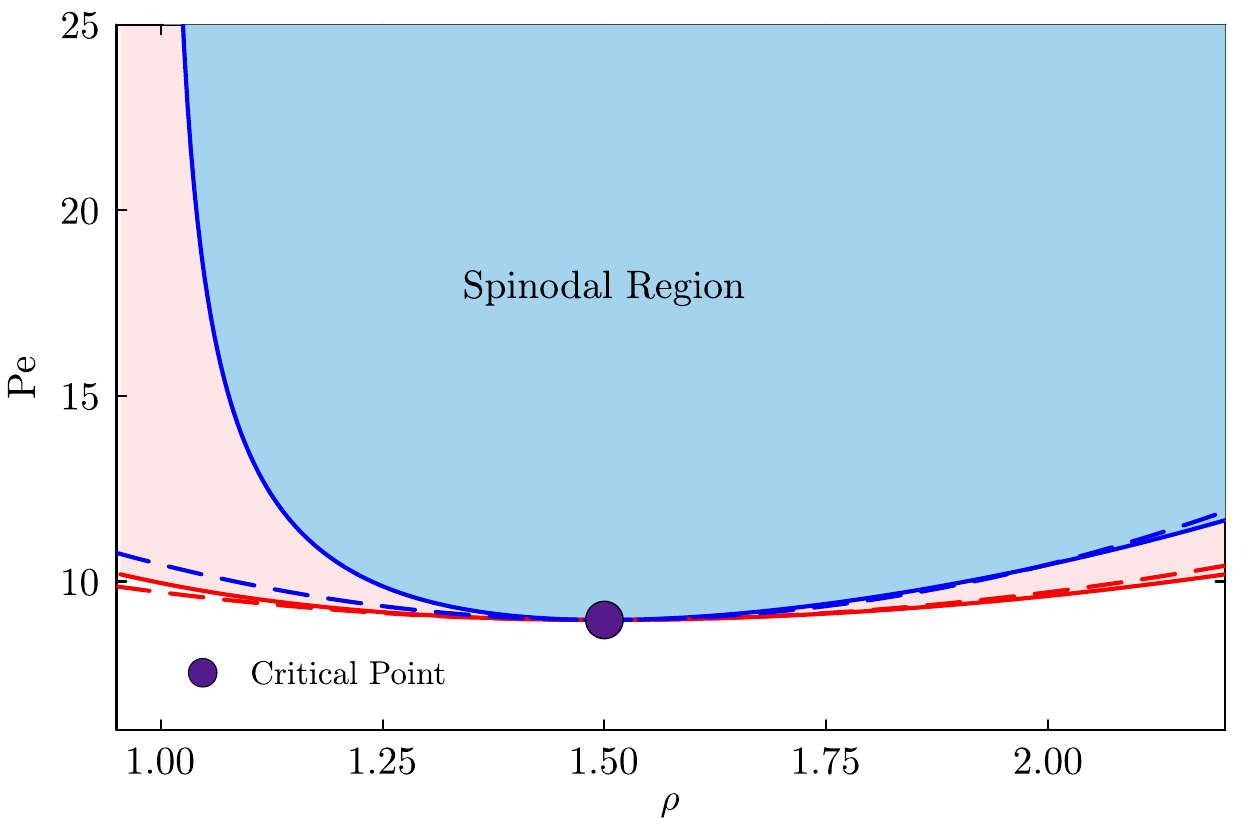}
    \caption{Phase diagram for quorum sensing MIPS. The solid blue line indicates the spinodal calculated from the hydrodynamic equations (\ref{eqn:deterministic mips 1}), above which homogeneous states are unstable and phase separate \cite{chaikin1995principles}. The red line is the binodal marking the phase separated densities, calculated using techniques from Ref. \cite{solon2018generalized}. The dashed blue and red lines mark the spinodal and binodal curves from the $\phi^4$ free energy (\ref{eqn: quorum free energy}), which agree with the exact curves near the critical point marked in purple. }
    \label{fig:mipsQuorumPhase}
\end{figure}

At deterministic level, the LDF (\ref{eqn:quorum action}) gives the hydrodynamic equations

\begin{equation}
    \begin{gathered}
        \pd_t\rho+\text{Pe}\pd_x (m e^{-\rho})=\pd_x^2 \rho,\\
        \pd_t m +\text{Pe}\pd_x (\rho e^{-\rho})=\pd_x ^2 m -2m,
    \end{gathered}
\label{eqn:quorum eqns}
\end{equation}
where $\text{Pe}$ is as in section \ref{sec:MIPS}, and we have absorbed $\sigma$ by re-scaling the density and magnetisation fields as $(\rho,m)\rar(\rho,m)/\sigma$. 

To locate the critical point we first calculate the spinodal curve using linear stability analysis. This results in the blue curve in \cref{fig:mipsQuorumPhase}, with the MIPS critical point at $\text{Pe}_c=2e^{3/2}$, $\rho_c=3/2$. Expanding the growth rate around this point we find similar behaviour to \cref{fig: Growth rate}, and so we can use the same slow space-time variables as in section \ref{sec:MIPS} to study the critical point. The binodals can also be determined exactly using the techniques of Ref. \cite{solon2018generalized}, with the calculation given in \ref{sec:appendix quorum binodals}, and the curves plotted as the solid red curves in \cref{fig:mipsQuorumPhase}.

As in the previous section, we now perform a weakly non-linear analysis on the critical point to extract an active model B equation, though, seeing as the calculational procedure is identical, we will just quote the result. To $O(r^3)$ away from the critical point we find an AMB equation
\begin{equation}
\pd_t\phi = \pd_x^2\left( \frac{\delta \mathcal{F}}{\delta\phi} - \frac{3}{4}(\pd_x\phi)^2 \right),
\label{eqn: Quorum AMB}
\end{equation}
where $\phi$ again is $\rho-\rho_c$, and the free energy is
\begin{equation}
\mathcal{F}=\int\dr\ -r\phi^2 + \frac{1}{6}\phi^4 + \frac{1}{4}(\pd_x\phi)^2.
\label{eqn: quorum free energy}
\end{equation}
As we found in the earlier AMB equation (\ref{eqn:ambp mips}), the $\lambda$ parameter is negative. This, however, is unsurprising, as other microscopic derivations of AMB based on quorum sensing also predict it to be negative \cite{cates2019active,solon2018generalized,dinelli2024fluctuating}. As shown in \cref{fig:mipsQuorumPhase}, the agreement between the binodal and spinodal curves from the free energy (\ref{eqn: quorum free energy}) and the full hydrodynamic equations is good near the critical point, however the free energy cannot capture the asymmetry in the full curves.

\subsection{Probability distribution}

Using the same approach on the full saddle equations we calculate the stationary probability distribution to be
\begin{equation}
\prob[\config]\asymp \exp\left[-\frac{2\sqrt{D}}{9\sigma\sqrt{\gamma}\ep}\mathcal{F}[\phi(x)]+O(r^2)\right],
\end{equation}
where the inverse factor of $\sigma$ comes from our rescaling the density and magnetisation fields. This result, along with that of the previous section, suggest, but by no means prove, that any one dimensional diffusive lattice gas undergoing the MIPS transition has near critical dynamics that live in the mean-field model B universality class.

\section{Flocking}
\label{sec: flocking}
\begin{figure}[t]
    \centering
    \begin{tikzpicture}[scale=0.65]
\draw[-,color=BlueViolet,line width= .8mm] (0,0) -- (10.6,0);
\filldraw[color=BlueViolet, fill=BlueViolet,thick] (-0.2,0) circle (.06);
\filldraw[color=BlueViolet, fill=BlueViolet,thick] (-0.5,0) circle (.06);
\filldraw[color=BlueViolet, fill=BlueViolet,thick] (-0.8,0) circle (.06);
\filldraw[color=BlueViolet, fill=BlueViolet,thick] (10.8,0) circle (.06);
\filldraw[color=BlueViolet, fill=BlueViolet,thick] (11.1,0) circle (.06);
\filldraw[color=BlueViolet, fill=BlueViolet,thick] (11.4,0) circle (.06);
\filldraw[color=Mahogany, fill=red!5,thick] (0.2,.4) circle (.2);

\filldraw[color=Blue, fill=blue!5,thick] (0.8,.4) circle (.2);
\filldraw[color=Blue, fill=blue!5,thick] (0.8,1) circle (.2);
\filldraw[color=Blue, fill=blue!5,thick] (1.4,.4) circle (.2);
\filldraw[color=Mahogany, fill=red!5,thick] (2,.4) circle (.2);
\filldraw[color=Mahogany, fill=red!5,thick] (2,1) circle (.2);
\filldraw[color=Blue, fill=blue!5,thick] (1.4,1) circle (.2);
\filldraw[color=Mahogany, fill=red!5,thick] 
(2.6,.4) circle (.2);
\filldraw[color=Blue, fill=blue!5,thick] 
(2.6,1) circle (.2);
\filldraw[color=Mahogany, fill=red!5,thick] 
(2.6,2.9) circle (.2);
\filldraw[color=Mahogany, fill=Red!5,thick] (3.2,.4) circle (.2);
\filldraw[color=Blue, fill=blue!5,thick] 
(3.2,1) circle (.2);
\filldraw[color=Blue, fill=blue!5,thick] 
(3.2,1.6) circle (.2);
\filldraw[color=Blue, fill=blue!5,thick] 
(4.4,.4) circle (.2);
\filldraw[color=Blue, fill=blue!5,thick] 
(4.4,1) circle (.2);
\filldraw[color=Blue, fill=blue!5,thick] 
(4.4,1.6) circle (.2);
\filldraw[color=Blue, fill=blue!5,thick] 
(3.8,.4) circle (.2);
\filldraw[color=Blue, fill=blue!5,thick] 
(3.8,1) circle (.2);
\filldraw[color=Blue, fill=blue!5,thick] 
(3.8,1.6) circle (.2);
\filldraw[color=Blue, fill=blue!5,thick] 
(3.8,2.2) circle (.2);
\filldraw[color=Blue, fill=blue!5,thick] (5,.4) circle (.2);
\filldraw[color=Blue, fill=blue!5,thick] (5,1) circle (.2);
\filldraw[color=Mahogany, fill=red!5,thick] (5,1.6) circle (.2);
\filldraw[color=Mahogany, fill=red!5,thick] (5.6,.4) circle (.2);
\filldraw[color=Blue, fill=blue!5,thick] (5.6,1) circle (.2);
\filldraw[color=Mahogany, fill=red!5,thick] (6.2,.4) circle (.2);
\filldraw[color=Mahogany, fill=red!5,thick] (6.2,1) circle (.2);
\filldraw[color=Mahogany, fill=red!5,thick] (6.8,.4) circle (.2);
\filldraw[color=Blue, fill=blue!5,thick] (7.4,.4) circle (.2);
\filldraw[color=Blue, fill=blue!5,thick] (8,.4) circle (.2);
\filldraw[color=Mahogany, fill=red!5,thick] (8.6,.4) circle (.2);
\filldraw[color=Mahogany, fill=red!5,thick] (9.8,.4) circle (.2);
\draw[-,color=BlueViolet,line width= .8mm] (0,-4) -- (10.6,-4);
\filldraw[color=BlueViolet, fill=BlueViolet,thick] (-0.2,-4) circle (.06);
\filldraw[color=BlueViolet, fill=BlueViolet,thick] (-0.5,-4) circle (.06);
\filldraw[color=BlueViolet, fill=BlueViolet,thick] (-0.8,-4) circle (.06);
\filldraw[color=BlueViolet, fill=BlueViolet,thick] (10.8,-4) circle (.06);
\filldraw[color=BlueViolet, fill=BlueViolet,thick] (11.1,-4) circle (.06);
\filldraw[color=BlueViolet, fill=BlueViolet,thick] (11.4,-4) circle (.06);
\draw [color=black!5,fill=black!5, smooth, samples=100, domain=0:10.6] plot(\x, {{sin(\x*25)*.3-2.5}}) -- plot[domain=10.6:0] (\x, {-4}) -- cycle; 
\draw [color=OliveGreen!5,fill=Green!20, smooth, samples=100, domain=0:10.6] plot(\x, {{sin(\x*20)*1.6-4}}) -- plot[domain=10.6:0] (\x, {-4}) -- cycle; 
\draw [color=OliveGreen, smooth, line width=0.3mm, samples=100, domain=0:10.6] plot(\x,{{sin(\x*20)*1.6-4}});
\draw [color=black, smooth, line width=0.3mm, samples=100, domain=0:10.6] plot(\x,{{sin(\x*25)*.3-2.5}});
\draw[-,color=BlueViolet,line width= .8mm] (0,-4) -- (10.6,-4);
\node[color=black] at (8.5,-2){$\rho(x)=\langle n^+ + n^-\rangle$};
\node[color=OliveGreen] at (2.3,-5){$m(x)=\langle n^+ - n^-\rangle$};
\draw[-to,color=Black, line width=.4mm] (9.8,.8) arc (130:70:.6);
\draw[-to,color=Black, line width=.4mm] (8.6,.8) arc (50:110:.6);
\draw[to-to,color=Black, line width=.4mm] (2.6,2) -- (2.6,2.6);
\filldraw[color=Blue, fill=blue!5,thick] (2.6,1.7) circle (.2);
\draw[-to,color=Black, line width=.4mm] (.4,.8) arc (50:110:.6);
\node at (0,1.5){$D+\lambda\ep$};
\node at (1.5,2.3){$\gamma\ep^2e^{-\beta}$};
\node at (8.6,1.4){$D+\lambda\ep$};
\node at (10,1.4){$D$};
\end{tikzpicture}
    \caption{Schematic of the flocking model. Multiple particles are allowed per site, and particles change species at a rate that depends on the local magnetisation. Blue particles are positive, red particles are negative. The first line gives the allowed microscopic moves. The second line shows the hydrodynamic fields, the gray gives the total density, and the green the magnetisation.}
    \label{fig:FlockingGas}
\end{figure}
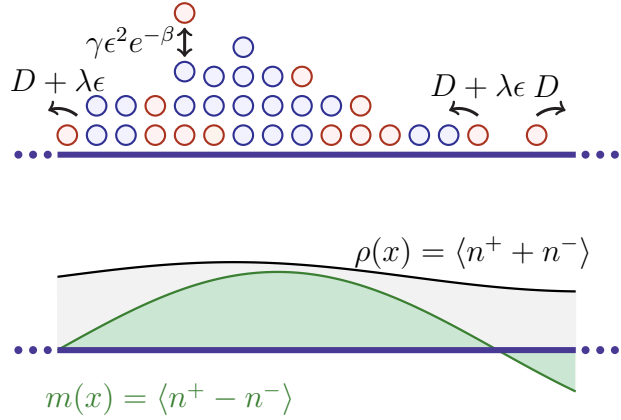

Our last active lattice gas is the one dimensional flocking model introduced in \cite{kourbane2018exact}, although similar to other lattice flocking models \cite{solon2013revisiting,solon2015flocking}. This model is designed to mimic the Vicsek mechanism for the transition to collective motion \cite{vicsek1995novel}. As in quorum sensing MIPS, we have two species of particle $n^{\pm}$ on a one dimensional lattice, with no limit to the number of particles on each site. Diffusion and self propulsion comes from plus (minus) particles hopping right (left) at rate $D+\lambda\ep$, and hopping left (right) at rate $D$. Alignment is induced through a Glauber type term, with $n^\pm$ particles switching type at rate $\gamma \ep^2\exp[\mp\beta(n^+-n^-)]$. See \cref{fig:FlockingGas} for a schematic.

As before, the Hamiltonian is written down using the general rules, giving

\begin{equation}
\begin{aligned}
\ham =&\sum_{i}\sum_{\alpha=\pm} Dn_i^{\alpha}\left(e^{\nh^\alpha_{i+1}-\nh^\alpha_i}+e^{\nh^\alpha_{i-1}-\nh^\alpha_i}-2\right)+\lambda\ep n^{\alpha}_i\left(e^{\nh^{\alpha}_{i+\alpha}-\nh^\alpha_i}-1\right)\\
&+\gamma\ep^2 n_i^\alpha e^{-\alpha\beta(n_i^+-n_i^-)}\left(e^{\nh^{-\alpha}_i-\nh^{\alpha}_i}-1\right).
\end{aligned}
\end{equation}
where the first term gives diffusion, the second self propulsion, and the the last gives flocking. 

Again, the rates are chosen such that diffusion dominates locally and controls the local equilibrium, which as in quorum sensing MIPS, is product Poissonian \cite{kourbane2018exact}. Performing the average over local equilibrium and Taylor expanding, we find
\begin{equation}
    \begin{aligned}
        \mathcal{A}=&\int\dr\int\wrt t\ \rh[\pd_t\rho+\lambda\pd_x m- D\pd_x^2 \rho]+\hat{m}[\pd_t m +\lambda\pd_x \rho-D\pd_x^2 m]\\
        &-D\rho(\pd_x\rh)^2-D\rho(\pd_x\hat{m})^2-2Dm(\pd_x \bar{\rho})(\pd_x \hat{m})-\\
        &\frac{\gamma e^{-\beta}}{2}(\rho+m)\exp[\rho(\cosh\beta-1)-m\sinh\beta]\left(e^{-2\hat{m}}-1\right)\\
        &-\frac{\gamma e^{-\beta}}{2}(\rho-m)\exp[\rho(\cosh\beta-1)+m\sinh\beta]\left(e^{2\hat{m}}-1\right),
    \end{aligned}
\end{equation}
where, as before, we have changed to density and magnetisation variables. This LDF has been found previously using Doi-Peliti theory, although it wasn't recognised as a large deviation function, being instead studied using renormalisation group \cite{scandolo2023active}.

\subsection{Deterministic behaviour}

The deterministic equations from this action agree with those derived in Ref. \cite{kourbane2018exact}, and are given by
\begin{equation}
    \begin{gathered}
        \pd_t \rho +\text{Pe}\pd_x m= \pd_x^2\rho,\\
        \pd_t m +\text{Pe}\pd_x\rho = \pd_x^2 m -2F(\rho,m),\\
    \begin{aligned}
F(\rho,m)= & (m \cosh [m \sinh (\beta)]-\rho \sinh [m \sinh (\beta)]) \\
& \times e^{-\beta+\rho \cosh (\beta)-\rho},
    \end{aligned}
    \end{gathered}
\label{eqn:flocking hydro determ}
\end{equation}
where we have performed the same re-scaling of space and time as in \cref{eqn:deterministic mips 1}. If the particles uniformly align, then the magnetisation becomes homogeneous and non-zero. Interestingly, this can happen even at zero self-propulsion ($\text{Pe}=0$), and homogeneous un-aligned states $\rho=\rho_0,m=0$ become unstable to aligned states when $\rho_0>(\sinh\beta)^{-1}$. A similar transition was studied numerically in the two-dimensional active Ising model and it was found to lie in the model A universality class \cite{solon2013revisiting,solon2015flocking}. Here we attempt to show this analytically.

To understand this transition we again perform a weakly non-linear analysis by  letting $\beta=\text{arcsinh}(1/\rho_0)(1+r)$ and expanding in $r$. Unlike the MIPS models, the relevant order parameter near the critical point, the magnetisation, is unconserved. This, along with a linear stability analysis of  (\ref{eqn:flocking hydro determ}), give the slow variables as $X=r^{1/2}x$, $\tau=r t$. As in (\ref{eqn: MIPS asymptotic expansion}) we expand the density and magnetisation in $r$ and substitute them into (\ref{eqn:flocking hydro determ}). To leading order we find  $\rho_1=0$, while at $O(r^{3/2})$ we have
\begin{equation}
\begin{gathered}
 \pd_T m_1 = \pd_X^2 m_1 +Am_1-6B\rho_0 m_1\rho_2 -2B m_1^3,\\
 B=\frac{1}{3\rho_0^2}\exp[-\rho_0+\sqrt{1+\rho_0^2} -\text{arcsinh}(1/\rho_0)].
\end{gathered}
\end{equation}
The presence of $\rho_2$ means we must work to higher order in $r$ to get a consistent set of equations, and at $O(r^2)$ we find
\begin{equation}
\begin{gathered}
        \pd_\tau\rho_2 = \pd_X^2 \rho_2,\\
    \pd_\tau m_2 = \pd_X^2 m_2 +Am_2-6B\rho_0 m_2\rho_2 -6B m_1^2 m_2,\\
    A=6B\rho_0\sqrt{1+\rho_0^2}\text{arcsinh}(1/\rho_0).
\end{gathered}
\end{equation}
Defining $\phi=r\rho_2$ and using the series expansion for $m$, the equations at $O(r^{3/2})$ and $O(r^2)$ can be summed into a single set, which upon returning to the original space-time variables are
\begin{subequations}
    \begin{gather}
        \pd_t \phi = \pd_x^2  \phi,\\
    \pd_t m = \pd_x^2 m +A rm-6B\rho_0 m\phi -2Bm^3.
    \end{gather}
\end{subequations}
These equations do not derive from a single free energy, and so initially one may think that the dynamics are not in an equilibrium class. However, the density perturbation is decoupled from the magnetisation and diffusively relaxes, while the magnetisation is still critically slowed. Therefore over long times we may assume that the density perturbation relaxes to zero. Substituting this into the magnetisation equation we find the dynamics of $m$ take model A form, with
\begin{equation}
    \pd_t m=-\frac{\delta \mathcal{F}}{\delta m},
\end{equation}
and a free energy
\begin{equation}
    \mathcal{F}=\int\dr\ -\frac{Ar}{2}m^2+\frac{B}{2}m^4+\frac{1}{2}(\pd_x m)^2.
\end{equation}
In other words, the free energy is still Ising type over long time scales, however the dynamics are different because there is no conservation law.


\section{Discussion}
\label{sec:discussion}
Starting from the master equations describing the stochastic dynamics we have derived hydrodynamic equations, and the corresponding large deviation functions describing the coarse grained behaviour of three active lattice gases. Using this hydrodynamic description we located their critical points and performed a series of weakly non-linear calculations, using them to argue that all models considered lay in mean-field kinetic Ising universality classes, model B for MIPS, and model A for flocking  \cite{hohenberg1977theory}.

These calculations were first done on the deterministic hydrodynamic equations, but were extended to include fluctuations by perturbatively solving the full saddle-point equations near the critical point. Doing this we found that the stationary probability distributions were given by the negative exponential of a free energy, just as in equilibrium, meaning that detailed balance is effectively restored asymptotically close to the critical point.

In the case of MIPS we also continued the expansion to higher order, showing that the two-field hydrodynamics reduced to active model B, the standard field theory of motility induced phase separation \cite{tjhung2018cluster,cates2019active}, as we left the critical region. This implies that non-equilibrium effects do become important as one departs from the critical point, however they are irrelevant at criticality. This agrees with a number of numerical studies on the MIPS critical point, of which most conclude that it is in the Ising universality class \cite{partridge2019critical,maggi2021universality,dittrich2021critical,speck2022critical}.

In this work we focused on one dimensional systems, but the methods used here could also be used in higher dimensional systems. Unfortunately, the LDFs become extremely cumbersome, as a new species must be introduced for every lattice direction. However, it is certainly worth investigating this further, as numerical studies of two dimensional lattice models of MIPS have  found some possible differences between models on square and triangular lattices \cite{speck2022critical,partridge2019critical}. Moreover, the active terms become more interesting in dimensions two or higher, with there being two independent non-equilibrium terms in the density current, rather than the one in one dimension \cite{tjhung2018cluster}. It would also be interesting to make these lattice gases thermodynamically consistent by explicitly accounting for the chemical fuel used in self-propulsion \cite{markovich2021thermodynamics,seifert2012stochastic}, which would give insight into the earlier phenomenological approaches which started with the continuum equations \cite{markovich2021thermodynamics}. We leave all this for future work.

\ack
We thank Tal Agranov and Tomohiro Sasamoto for helpful discussions, and Tanniemola Liverpool for a careful reading of the manuscript. This work was supported by an EPSRC studentship.

\appendix
  
\section*{Appendices}
\section{Path integral derivation}
\label{sec:appendix path derivation}
The path integral representation is derived here for single species on one site, but the derivation is identical for multiple sites. To put the master equation in path integral form we use linearity to evolve the probability forwards by a small amount of time $\delta t$, $\prob_n(\delta t)=\exp[\ham\delta t]\prob_n(0)$.
\begin{align*}
    e^{\ham\delta t}\prob_n(0)& \approx [1+\ham\delta t]\prob_n(0)\\
    &= [1+\delta t(e^{-\Delta n\pd_n}-1)r(n)]\prob_n(0)\\
    &= \sum_{n_0} [1+\delta t(e^{-\Delta n_0 \pd_n}-1)\delta_{n,{n_0}}r(n_0)]\prob_{n_0}(0)\\
    &= \sum_{n_0}\int_{-\ci\pi}^{\ci\pi}\frac{\wrt \nh_0}{2\pi \ci}[1+\delta t(e^{-\Delta n_0 \pd_n}-1)e^{-\nh_0(n-n_0)}r(n_0)]\prob_{n_0}(0)\\
    &= \sum_{n_0}\int_{-\ci\pi}^{\ci\pi}\frac{\wrt \nh_0}{2\pi \ci}[1+\delta t(e^{\Delta n_0 \nh_0}-1)e^{-\nh_0(n-n_0)}r(n_0)]\prob_{n_0}(0)\\
    &\approx \sum_{n_0}\int_{-\ci\pi}^{\ci\pi}\frac{\wrt \nh_0}{2\pi \ci}\exp[-\nh_0(n-n_0)+\delta t(e^{\nh_0\Delta n}-1)r(n_0) ]\prob_{n_0}(0),
    \label{eqn: discrete path derivation}
\end{align*}
where in the last line we have re-exponentiated the result, which is correct to $O(\delta t)$ \cite{negele2018quantum}. By combining many of these evolutions we get the probability at a time $T$
\begin{subequations}
    \begin{gather}
        \prob_n(T)=\int \mathcal{D}[n,\nh] e^{-\mathcal{S}}\prob_n(0),\\
        \mathcal{S}=\sum_{t=0}^{t=T-\delta t} \nh_t\left(n_{t+\delta t}-n_t\right)-\delta t\left(e^{\Delta n \nh_t}-1\right) r\left(n_t\right),\\
        \int\mathcal{D}[n,\nh]=\lim _{\delta t \rightarrow 0} \prod_{t=0}^{T-\delta  t} \sum_{n(t)=0}^{N_{max}} \int_{-\ci \pi}^{\ci \pi} \frac{\mathrm{d} \nh(t)}{2 \pi \ci}.
    \end{gather}
\end{subequations}
It is tempting to take the continuum limit and convert the forward differences to time derivatives.  However, $n$ is always an integer valued quantity and a derivative does not exist. With this fact in mind we will write the path integral in continuous notation although it should be understood in its discrete form. In continuum form, the action is
\begin{equation}
    \mathcal{S}=\int\wrt t\ \nh\pd_t n -\ham(\nh,n),
\end{equation}
which is the standard form from classical mechanics if $\nh$ is the conjugate `momentum' to $n$, obeying canonical commutation relations $[n,\hat{n}]=1$. The way we have discretised time also means that the continuum Langevin equations from this integral must be interpreted in the It\^{o} sense \cite{cates2022stochastic}.

Using this path integral the probability of a single trajectory through configuration space is given by
\begin{equation}
\prob[\text{traj}]=\int\mathcal{D}[\hat{n}]e^{-\mathcal{S}}\prob(t=0),
\end{equation}
where only $\hat{n}$ is integrated over, while the value of $n$ is specified at each time step.

\section{Current form of the LDF}
\label{sec:current LDF}

Macroscopic fluctuation theory considers the joint fluctuations of the density $\rho$, conserved current $j$, and flux $K$ \cite{bertini2015macroscopic}, however our LDFs are written in terms of the conjugate field $\rh$. To demonstrate how to convert between the two we use the following LDF describing diffusion combined with particle creation and destruction at rates $b$ and $d$ respectively
\begin{equation}
\mathcal{A}=\int\wrt t\int\wrt x\ \rh[\pd_t \rho-D\pd_x^2\rho]-D\rho(\pd_x \rh)^2-b(e^{\rh}-1)-d(e^{-\rh}-1)\rho.
\label{eqn:supp react}
\end{equation}
To put this in current form we demand that the density satisfies the exact equation
\begin{equation}
\pd_t\rho=-\pd_x j +K,
\label{eqn: appen continuity}
\end{equation}
where $\rho$, $j$, and $K$ are fluctuating quantities representing the density, conserved, and un-conserved currents. Let us now break apart the LDF into two pieces coming from the conserved and non-conserved current. For the conserved current alone we have an 
\begin{equation}
\mathcal{A}=\int\wrt t\int\wrt x\ \rh[\pd_t{\rho}-D\pd_x^2\rho]-D\rho(\pd_x \rh)^2.
\end{equation}
We now replace $\pd_t\rho$ with $-\pd_x j$ and integrate by parts, giving the result
\begin{equation}
\mathcal{A}_j=\int\wrt t\int\wrt x\ j\pd_x \rh+D(\pd_x\rho)(\pd_x \rh)-D\rho(\pd_x \rh)^2.
\end{equation}
We would now like to perform the path integral over $\rh$, and this can be done exactly, giving
\begin{equation}
\mathcal{A}_j=\int\wrt t\int\wrt x\ \frac{(j+D\pd_x\rho)^2}{4D\rho}.
\end{equation}
This quantifies how the conserved current $j$ fluctuates around it's average value $-D\pd_x \rho$. We now turn to the non-conserved current, with an LDF
\begin{equation}
\mathcal{A}=\int\wrt t\int\wrt x\ \rh\pd_t {\rho}-b(e^{\rh}-1)-d(e^{-\rh}-1)\rho.
\end{equation}
Following similar methods to before we replace $\pd_t\rho$ with $K$ and perform the path integral over $\rh$. The $\rh$ path integral cannot be performed exactly but can be done using a saddle point method. The result is that $\rh$ is given by
\begin{equation}
e^{\rh}=\frac{K+\sqrt{K^2+4db\rho}}{2b}.
\end{equation} 
This is substituted into the LDF to give 
\begin{equation}
\mathcal{A}_K=\int\wrt t\int\wrt x\ b+d\rho-\sqrt{K^2+4bd\rho}+K\log\frac{\sqrt{K^2+4bd\rho}+K}{2b}.
\end{equation}
The total probability for a given path is then
\begin{equation}
\prob\asymp \delta(\pd_t\rho+\pd_xj-K)e^{-\mathcal{I}(\rho,j,K)/\ep}
\end{equation}
where the large deviation functional is
\begin{equation}
\mathcal{I}(\rho,j,K)=\mathcal{A}_j+\mathcal{A}_K,
\end{equation}
and the functional delta function ensures that the exact equation (\ref{eqn: appen continuity}) is satisfied. The LDF in terms of $\rho$ and $\rh$ can be recovered by introducing a functional representation of the delta function and extremising the action over $K$ and $j$ \cite{bertini2015macroscopic}.

\section{MIPS LDF}
\label{sec:appendix mips derivation}
Here we explicitly derive the hydrodynamic action for the MIPS model. Starting with \cref{eqn: discrete mips action} we first perform the sum over $\alpha$ and then take the average over the local equilibrium measure to get
\begin{equation}
    \begin{aligned}
        \ham &=\sum_i D\rho^+_i(1-\rho_{i+1})\left(e^{\rh^+_{i+1}-\rh^+_{i}}-1 \right)+D\rho^+_i(1-\rho_{i-1})\left(e^{\rh^+_{i-1}-\rh^+_{i}}-1 \right)\\
        &+D\rho^-_i(1-\rho_{i+1})\left(e^{\rh^-_{i+1}-\rh^-_{i}}-1 \right)+D\rho^-_i(1-\rho_{i-1})\left(e^{\rh^-_{i-1}-\rh^-_{i}}-1 \right)\\
        &+D\rho^+_i\rho^-_{i-1}\left(e^{\rh^+_{i-1}-\rh^+_i -\rh^-_{i-1}+\rh^-_i}-1\right)+D\rho^+_i\rho^-_{i+1}\left(e^{\rh^+_{i+1}-\rh^+_i -\rh^-_{i+1}+\rh^-_i}-1\right)\\
        &+\lambda\ep \rho^+_i(1-\rho_{i+1})\left(e^{\rh^+_{i+1}-\rh^+_{i}}-1\right)+\lambda\ep \rho^-_i(1-\rho_{i-1})\left(e^{\rh^-_{i-1}-\rh^-_{i}}-1\right)\\
        & +\gamma\ep^2 \rho^+_i \left(e^{\rh^-_i-\rh_i^+}-1\right)+\gamma\ep^2 \rho^-_i \left(e^{\rh^+_i-\rh_i^-}-1\right).
    \end{aligned}
\end{equation}
Exchanging the sum for an integral and Taylor expanding the fields we find
\begin{equation}
    \begin{split}
        \ham &=\ep \int\dr\ (\pd_x\rh^+)^2\rho^+(1-\rho^+)+(\pd_x\rh^-)^2\rho^-(1-\rho^-)-2(\pd_x\rh^+)(\pd_x\rh^-)\rho^+\rho^-\\
        &-2(\pd_x\rh^-)(\rho^+\pd_x\rho^-+\rho^-\pd_x\rho)-2(\pd_x\rh^+)(\pd_x\rho^+)\rho^+ +(\pd_x^2\rh^-)(\rho^-(\rho^-)^2-2\rho^+\rho^-)\\ 
        &+(\pd_x^2\rh^+)\rho^+(1-\rho^+)
        \lambda \rho^+(1-\rho)(\pd_x\rh^+)-\lambda \rho^+(1-\rho)(\pd_x\rh^+)+\gamma \rho^+ \left(e^{\rh^- -\rh^+}-1\right)\\
        &+\gamma\rho^- \left(e^{\rh^+ -\rh^-}-1\right),
    \end{split}
\end{equation}
where $\rho=\rho^+ +\rho^-$. After changing to density and magnetisation variables: $\rho=\rho^{+}+\rho^{-}$, $m=\rho^{+}-\rho^{-}$, $\rh^+=(\rh+\hat{m})$, $\rh^-=(\rh-\hat{m})$ and performing many integrations by parts we arrive at the LDF from the main text.

\section{Exact binodals for Quorum sensing MIPS}
\label{sec:appendix quorum binodals}

To calculate the exact binodals for the Quorum sensing model of MIPS we follow the approach of Ref. \cite{solon2018generalized}. We begin by setting both time derivatives in the deterministic equations (\ref{eqn:quorum eqns}) to zero. Solving the magnetisation equation gives
\begin{equation}
m = \frac{1}{\text{Pe}}\pd_x e^{\rho},
\end{equation}
which is then substituted into the density equation, giving
\begin{equation}
\pd_x g=0,\quad g = \text{Pe}\rho e^{-\rho} +\frac{2}{\text{Pe}} e^{\rho} -\frac{e^{\rho}}{\text{Pe}}(\pd_x\rho)^2 -\frac{e^{\rho}}{\text{Pe}}\pd_x^2\rho.
\label{eqn:binodal g current}
\end{equation}
This current $g$ takes the form of an ordinary AMB current and so we can calculate the binodals using the pseudo-variables of Ref. \cite{solon2018generalized}. These are defined as
\begin{equation}
R=e^{\rho},\quad \varphi = \frac{R^2}{\text{Pe}}+\frac{1}{2}\text{Pe}(\log R)^2,
\end{equation}
and denoting the non-derivative terms in (\ref{eqn:binodal g current}) as $g_0$, the binodal densities have equal values of $\varphi'(R) R-\varphi$, and $g_0$. These values can be found numerically using Newton's method, and the corresponding densities yield the binodal curve of \cref{fig:mipsQuorumPhase}.

\section*{References}
\bibliographystyle{unsrt}
\bibliography{biblio}

\end{document}